# A Neural Network-enhanced Reproducing Kernel Particle Method for Modeling Strain Localization

Jonghyuk Baek[1], J. S. Chen[1*], and Kristen Susuki[1]

[1]University of California, San Diego

## Abstract

Modeling the localized intensive deformation in a damaged solid requires highly refined discretization for accurate prediction, which significantly increases the computational cost. Although adaptive model refinement can be employed for enhanced effectiveness, it is cumbersome for the traditional mesh-based methods to perform while modeling the evolving localizations. In this work, neural network-enhanced reproducing kernel particle method (NN-RKPM) is proposed, where the location, orientation, and shape of the solution transition near a localization is automatically captured by the NN approximation via a block-level neural network optimization. The weights and biases in the blocked parametrization network control the location and orientation of the localization. The designed basic four-kernel NN block is capable of capturing a triple junction or a quadruple junction topological pattern, while more complicated localization topological patters are captured by the superposition of multiple four-kernel NN blocks. The standard RK approximation is then utilized to approximate the smooth part of the solution, which permits a much coarser discretization than the high-resolution discretization needed to capture sharp solution transitions with the conventional methods. A regularization of the neural network approximation is additionally introduced for discretization-independent material responses. The effectiveness of the proposed NN-RKPM is verified by a series of numerical verifications.

# 1. Introduction

Modeling many complicated and evolving localizations, such as, fracture and shear bands, is a challenging task. In fracture modeling, significant developments have been made over the past several decades, such as discrete crack approaches and diffuse crack approaches. The discrete crack approaches include meshfree methods with visibility/diffraction criterion[1,2], local partition of unity-based enrichment[3,4], and extended and generalized finite element methods[5–7]. This class of methods can yield mesh-independent solutions but requires surface tracking, which becomes ineffective when the crack path is highly complicated. The diffuse crack approaches offer the possibility of modeling complicated crack patterns without surface tracking. However, the solution of this method-type is prone to being discretization-sensitive. If the discretization sensitivity is not properly addressed, numerical results including energy, force, and damage patterns can be strongly affected by the choice of discretization due to the loss of ellipticity in the mathematical formulation[8]. For the diffuse crack approaches, nonlocal models with intrinsic length scales, such as quantity averaging[9], high order gradient models[10–12], and phase field methods[13–16] have successfully remedied the discretization sensitivity. However, nonlocal methods often require very fine discretization, which significantly increases the computational cost. While adaptive model refinement can be employed for enhanced effectiveness, it is cumbersome for the traditional mesh-based methods to perform.

The deep artificial neural network has demonstrated its ability in information processing, such as image recognition, and is rapidly extending its regime to various fields, including mechanistic machine learning. Its successful widespread adoption is mainly attributed to the adaptive nature of manipulating the function space based on data via minimization of the loss function. Such flexibility allows a computational model to capture certain information that is hard to identify by conventional approaches. For example, He et al. (2021) [17] introduced the deep autoencoder to discover the latent low-dimensional data pattern embedded in noisy material datasets, which significantly enhanced the effectiveness of the physics-constrained data-driven approach developed by He and Chen (2020)[18] and He et al. (2021)[19]. Lee and Carlberg (2020) [20] proposed a model order reduction technique based on low-dimensional nonlinear manifolds constructed by the deep convolutional autoencoder and demonstrated the enhanced performance over linear subspace-based reduced order modeling techniques.

Recently, the utilization of neural networks (NNs) for solving partial differential equations (PDE) in physics and mechanics has gained increased attention. This class of studies explores the NN's ability to be used as an approximation, considering its flexible function space construction. Raissi et al. (2019) [21] proposed a physics-informed neural network (PINN), which serves as an approximation of the solution of a PDE in the framework of a collocation method, and showed the effectiveness of the method on the one-dimensional Burgers equation using densely-connected, deep neural networks with the hyperbolic tangent activation function. Haghighat and Juanes (2021) [22] introduced a Python package SciANN for scientific computing using PINN and showed that the method captured strain and stress localization produced in a perfectly plastic material. However, they used 4 densely-connected layers with 100 neurons per layer, which involves 100 million unknowns, to obtain the results. Samaniego et al. (2020) [23] showed that the potential-based loss functions lead to better results with significantly less unknowns than the collocation-based loss function that is widely used in PINN. However, the aforementioned studies employed multiple fully-connected hidden layers, which may not be an optimal choice to the solution of PDEs. Zhang et al. (2020) [24] devised the deep neural network to represent standard approximations such as finite elements and reproducing kernel particles, and treated the nodal positions as unknown network parameters. This element- or particle-wise network is a sparse network, indicating a potential of designing a more efficient network. Although NN for solving PDEs in physics and mechanics has drawn increased attention, the research on development of NN for modeling discontinuities and localizations remains very limited and deserves further investigation.

While a neural network-based approximation can be highly effective in capturing localizations due to the adaptive nature of the function space, it may be inefficient to employ the NN-based approximation across the entire domain, in which the vast majority exhibits a smooth solution that can be accurately approximated by traditional methods, such as meshfree methods[25,26] and isogeometric analysis[27]. Particularly, in the reproducing kernel (RK) approximation[25,26], the order of continuity of the approximation and the order of basis are independently controlled, and the RK approximation has been proven to effectively and accurately capture smooth solutions with a coarse domain discretization. In this work, the RK approximation will be utilized to approximate the smooth part of the solutions while a NN approximation will be NN designed to effectively capture sharp solution transitions with proper regularization for dissipation energy objectivity.

The paper is organized as follows. In Section 2, the basic equations will be provided, including the minimization problems under consideration and the reproducing kernel approximation. In Section 3, a neural network-enhanced reproducing kernel approximation is presented. In Section 4, the implementation details including the neural network structure and solution procedure are provided. This is followed by numerical examples in Section 5 and concluding remarks in Section 6.

# 2. Basic Equations

## 2.1. Minimization problems under consideration

Let $\Omega \in \mathbb{R}^d$ be a domain with the space dimension $d$ and $\partial \Omega = \partial \Omega_g \cup \partial \Omega_h$ be the boundary of $\Omega$ with the Dirichlet boundary $\partial \Omega_g$ and the Neumann boundary $\partial \Omega_h$. In this work, the following minimization problem is considered:

$$\min_{\mathbf{u}} \Pi(\mathbf{u}) = \int_{\Omega} \psi(\mathbf{u}) \, d\Omega - F(\mathbf{u}) + \frac{\alpha}{2} \int_{\partial \Omega_g} (\mathbf{u} - \overline{\mathbf{u}}) \cdot (\mathbf{u} - \overline{\mathbf{u}}) \, d\Omega, \tag{1}$$

where $\mathbf{u}$ is the displacement vector, $\psi(\mathbf{u})$ is the strain energy density functional, and $F(\mathbf{u})$ is the external energy functional. The Dirichlet boundary condition $\mathbf{u} = \overline{\mathbf{u}}$ on $\partial \Omega_g$ is imposed by the penalty method with the penalty parameter $\alpha$. For linear elasticity, $\psi(\mathbf{u}) = \psi^{el}(\mathbf{u})$ and $F(\mathbf{u})$ are taken to have the following form:

$$\psi^{el}(\mathbf{u}) = \frac{1}{2} \boldsymbol{\varepsilon}(\mathbf{u}) : \mathbf{C} : \boldsymbol{\varepsilon}(\mathbf{u})$$

$$F(\mathbf{u}) = \int_{\Omega} \mathbf{u} \cdot \mathbf{b} \, d\Omega + \int_{\partial \Omega_h} \mathbf{u} \cdot \mathbf{h} \, d\Gamma \tag{2}$$

where $\boldsymbol{\varepsilon}$, $\mathbf{C}$, $\mathbf{b}$, and $\mathbf{h}$ are the strain tensor, elasticity tensor, body force vector, and boundary traction vector, respectively. For tension-compression decomposed elasticity-damage problems, $\psi(\mathbf{u}) = \psi^D(\mathbf{u})$ are defined as follows[13]:

$$\psi^D(\mathbf{u}) = g\left(\eta(\boldsymbol{\varepsilon}(\mathbf{u}))\right)\psi_0^+(\mathbf{u}) + \psi_0^-(\mathbf{u}) + \bar{\psi}\left(\eta(\boldsymbol{\varepsilon}(\mathbf{u}))\right) \tag{3}$$

where the damage parameter, $\eta$, degradation function, $g(\eta)$, and dissipation energy density functional, $\bar{\psi}$, are introduced, and $\psi_0^+$ and $\psi_0^-$ are tensile and compressive strain energies, respectively, denoted as

$$\psi_0^+ = \mu\langle\bar{\varepsilon}_i\rangle\langle\bar{\varepsilon}_i\rangle + \frac{\lambda}{2}\langle\sum\bar{\varepsilon}_i\rangle^2$$
$$\psi_0^- = \psi^{el} - \psi_0^+ \tag{4}$$

with the principal strain tensor, $\bar{\boldsymbol{\varepsilon}}$, Lamé's first and second parameters, $\mu$ and $\lambda$, and Macaulay brackets, $\langle\cdot\rangle = \max(\cdot,0)$.

In this work, the following damage model is employed:

$$\eta(\kappa) = \min\left(1, \max\left(0, \bar{\eta}(\kappa)\right)\right) \tag{5}$$

$$\bar{\eta}(\kappa) = \frac{1 - \kappa_0/\kappa}{1 - \kappa_0/\kappa_c} \tag{6}$$

$$g(\eta) = 1 - \eta \tag{7}$$

where $\kappa$, $\kappa_0$, and $\kappa_c$ are irreversible internal state variable, limit elastic strain, and critical equivalent strain, respectively, with $\kappa$ satisfing the following Kuhn-Tucker conditions:

$$\dot{\kappa} \geq 0$$

$$\bar{\kappa} - \kappa \leq 0 \tag{8}$$

$$\dot{\kappa}(\bar{\kappa} - \kappa) = 0$$

where $\bar{\kappa}$ is defined as

$$\bar{\kappa} = \sqrt{\frac{2\psi_0^+}{E}}. \tag{9}$$

Note that, in one-dimensional uniaxial tension, $\bar{\kappa} = \varepsilon_{11}$. For this considered damage model, $\bar{\psi}$ in (3) is defined as follows:

$$\bar{\psi}(\eta) = p\left(\frac{1}{q - \eta} - \frac{1}{q}\right) \tag{10}$$

where $p$ and $q$ are as follows:

$$p = \frac{E}{2}(\kappa_0 q)^2, \quad q = \frac{\kappa_c}{\kappa_c - \kappa_0}. \tag{11}$$

The stress $\boldsymbol{\sigma}$ defined as

$$\boldsymbol{\sigma} = \left(1 - \eta(\boldsymbol{\varepsilon}(\mathbf{u}))\right)\frac{\partial \psi_0^+(\mathbf{u})}{\partial \boldsymbol{\varepsilon}(\mathbf{u})} + \frac{\partial \psi_0^-(\mathbf{u})}{\partial \boldsymbol{\varepsilon}(\mathbf{u})}. \tag{12}$$

The dissipation functional defined in Eq. (10) is derived such that, for the adopted damage law in Eq. (5)-(7), the strain energy density functional $\psi^D$ satisfies $\psi^D = \int \boldsymbol{\sigma} : d\boldsymbol{\varepsilon}$, which leads to $\bar{\psi} = \int \psi_0^+ d\eta$. Note that with the dissipation functional (10), the minimization problem in (1) is consistent with the standard variational formulation shown below:

$$\int_\Omega \delta\boldsymbol{\varepsilon}(\mathbf{u}) : \boldsymbol{\sigma}(\boldsymbol{\varepsilon}) \, d\Omega = \int_\Omega \delta\mathbf{u} \cdot \mathbf{b} \, d\Omega + \int_{\partial\Omega^h} \delta\mathbf{u} \cdot \mathbf{h} \, d\Gamma. \tag{13}$$

In this work, $\Pi(\mathbf{u})$ in Eq. (1) is taken as the loss function for the numerical procedures in the neural network framework. The dissipation functional $\bar{\psi}$ introduces additional energy to the elastic energy, $g\big(\eta(\boldsymbol{\varepsilon}(\mathbf{u}))\big)\psi_0^+(\mathbf{u}) + \psi_0^-(\mathbf{u})$, such that the stationary condition of the potential energy functional $\delta\Pi(\mathbf{u}) = 0$ yields the variational equation in (13) with the desired stress degradation in (12).

### 2.2. Reproducing kernel approximation

In this work, the standard reproducing kernel (RK) approximation[25,26] for solving the minimization problem will be enriched with a neural network (NN) approximation (see Section 3) to better capture the strain localization. In the proposed approach, the RK approximation is used to approximate the smooth part of the solution while the NN approximation is used to approximate the sharp solution transition near localization. In this subsection, the RK approximation is introduced as the basis for the smooth part of the solution; the NN enhancement will be discussed in the next section.

Let us consider a domain $\Omega$ discretized by $NP$ nodes, as shown in Figure 1, with nodal coordinate $\mathbf{x}_J$ with $1 \leq J \leq NP$. The RK approximation, $u^h(\mathbf{x})$, of a function $u(\mathbf{x})$ is

$$u^h(\mathbf{x}) = \sum_{J=1}^{NP} \Psi_J(\mathbf{x}) d_J, \tag{14}$$

where $\Psi_J(\mathbf{x})$ is RK shape function of node $J$ and $d_J$ is the generalized nodal coefficient of node $J$. The RK shape function, $\Psi_J(\mathbf{x})$, is a correction of a kernel function, $\Phi_a(\mathbf{x} - \mathbf{x}_J)$, defined on the compact support of node $J$ with a support size of $a$:

$$\Psi_J(\mathbf{x}) = \left\{ \sum_{|\boldsymbol{\alpha}| \leq n} (\mathbf{x} - \mathbf{x}_J)^{\boldsymbol{\alpha}} b_{\boldsymbol{\alpha}}(\mathbf{x}) \right\} \Phi_a(\mathbf{x} - \mathbf{x}_J), \tag{15}$$

where $(\mathbf{x} - \mathbf{x}_J)^{\boldsymbol{\alpha}}$ is a basis function, $\boldsymbol{\alpha} = (\alpha_1, \alpha_2, \ldots, \alpha_d)$ with dimension $d$, and $|\boldsymbol{\alpha}| \equiv \sum_{i=1}^{d} \alpha_i$. $\mathbf{x}^{\boldsymbol{\alpha}}$ is defined as

$$\mathbf{x}^{\boldsymbol{\alpha}} \equiv x_1^{\alpha_1} \cdot x_2^{\alpha_2} \cdot \ldots \cdot x_d^{\alpha_d}. \tag{16}$$

The coefficients, $b_{\boldsymbol{\alpha}}(\mathbf{x})$ for $|\boldsymbol{\alpha}| \leq n$, are the solutions of the following set of reproducing conditions:

$$\sum_{J=1}^{NP} \Psi_J(\mathbf{x}) \mathbf{x}_J^{\boldsymbol{\alpha}} = \mathbf{x}^{\boldsymbol{\alpha}}, \qquad |\boldsymbol{\alpha}| \leq n, \tag{17}$$

which leads to the explicit form of $\Psi_J(\mathbf{x})$ as follows.

$$\Psi_J(\mathbf{x}) = \mathbf{H}^T(\mathbf{0})\mathbf{M}^{-1}(\mathbf{x})\mathbf{H}(\mathbf{x} - \mathbf{x}_J)\Phi_a(\mathbf{x} - \mathbf{x}_J), \tag{18}$$

where $\mathbf{M}(\mathbf{x})$ is the moment matrix and $\mathbf{H}(\mathbf{x} - \mathbf{x}_J)$ is the basis vector defined as

$$\mathbf{M}(\mathbf{x}) = \sum_{J=1}^{NP} \mathbf{H}(\mathbf{x} - \mathbf{x}_J)\mathbf{H}^T(\mathbf{x} - \mathbf{x}_J)\Phi_a(\mathbf{x} - \mathbf{x}_J), \tag{19}$$

$$\mathbf{H}(\mathbf{x} - \mathbf{x}_J) = \left[1, (x_1 - x_{1J}), (x_2 - x_{2J}), (x_3 - x_{3J}), \cdots, (x_3 - x_{3J})^n\right]^T. \tag{20}$$

The order of continuity of the RK approximation is determined by the kernel function, $\Phi_a(\mathbf{x} - \mathbf{x}_J)$, while the polynomial completeness of the approximation is determined by the basis vector, $\mathbf{H}(\mathbf{x} - \mathbf{x}_J)$. Thus, it is straightforward to introduce high order continuity into the approximation space, which makes the RK approximation more appealing for approximating the smooth part of solution than the $C^0$ interpolation-type approximations used in finite element methods.

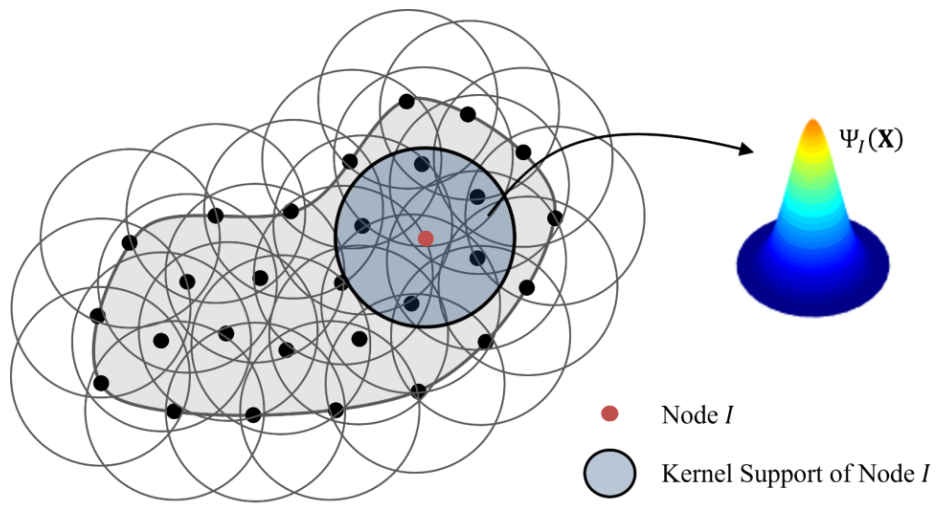

Figure 1. Illustration of RK discretization and shape function of node *I*

## 3. Neural Network-enhanced Reproducing Kernel Approximation

In this chapter, a neural network-enhanced reproducing kernel (NN-RK) approximation is proposed. Neural network (NN) architectures are developed with interpretable weights and biases to automatically detect the locations and orientations of strain localization, and to construct functions with sharp transitions near the localizations. This section presents the details of the network structure that incorporates the RK approximation.

Let the approximation $u^h(\mathbf{x}) \approx u(\mathbf{x})$ be decomposed into reproducing kernel (RK) and neural network (NN) approximations.

$$u^h(\mathbf{x}) = u^{RK}(\mathbf{x}) + u^{NN}(\mathbf{x}), \tag{21}$$

where $u^{RK}(\mathbf{x})$ and $u^{NN}(\mathbf{x})$ are the RK approximation and NN approximation, respectively. In the above approximation, the NN approximation is constructed to detect localization and to introduce a fine-scale feature to the solution in a region close to a localization, while the RK approximation is employed to capture the overall smooth behavior of the solution. The RK approximation takes its traditional form discussed in Section 2:

$$u^{RK}(\mathbf{x}) = \sum_{J=1}^{NP} \Psi_J(\mathbf{x}) d_J, \tag{22}$$

where $\Psi_J(\mathbf{x})$ and $d_J$ are RK shape function and RK generalized coefficient associated with RK node $J$, respectively.

## 3.1. Considerations on the construction of neural network approximation

In the construction of the NN approximation, we focus on the following considerations: 1) the *sharp solution transition associated with highly localized strain with weak discontinuity* be approximated by NN approximation, 2) the position and orientation of *complex localization paths* be automatically captured, 3) the NN approximation influences only *local regions* close to localizations.

In Consideration 1, the approximation of weak discontinuities in the displacement field on the interface between damaged and undamaged materials is considered, as illustrated in Figure 2.

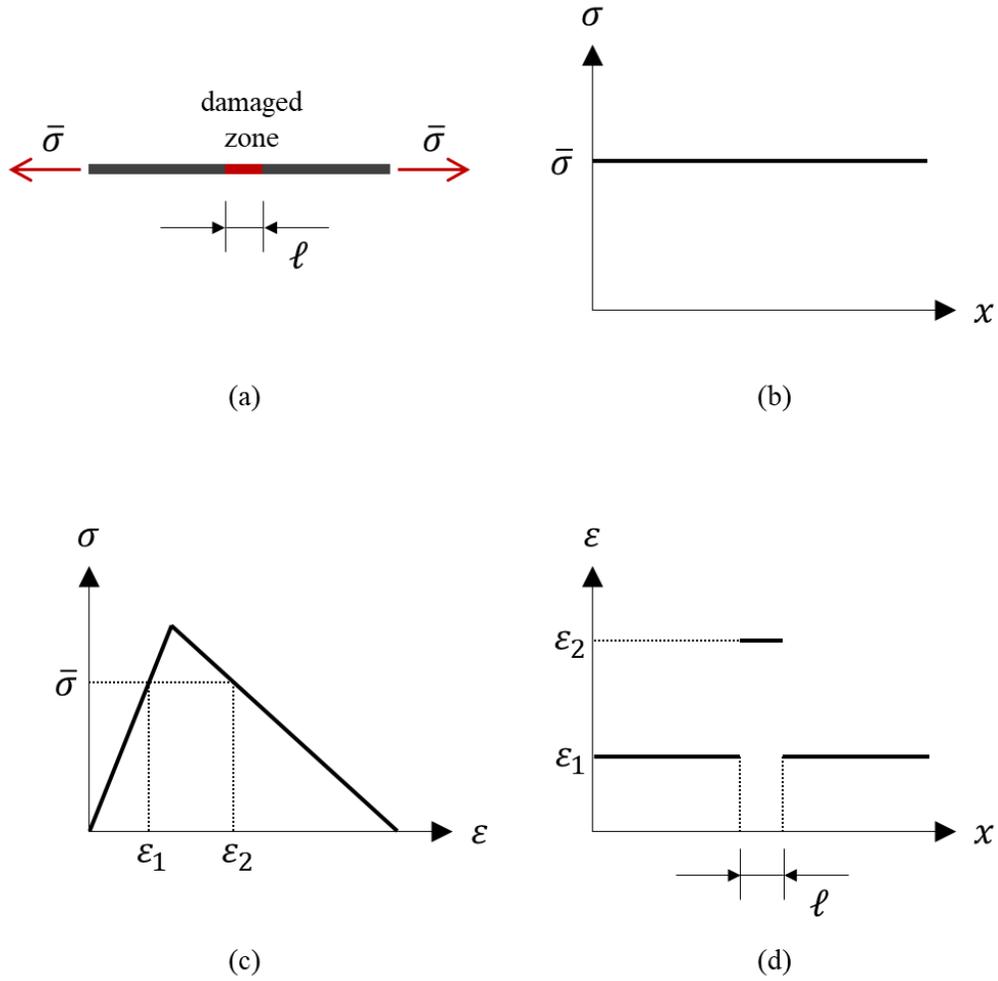

Figure 2. Illustration of discontinuous strain field in a damaged material under tensile stress: (a) a bar with a damaged zone of width, $\ell$, under tensile stress of $\bar{\sigma}$, (b) uniform stress in equilibrium, (c) two strain values corresponding to $\bar{\sigma}$, and (d) strain field with a discontinuity

To achieve Consideration 2, unknown parameters representing the localization position and orientation will be treated as unknowns and be included in the NN approximation. Also, a block-level NN approximation will be constructed to capture complex localization topology by a superposition of multiple block-level NN approximations, each of which are constructed to approximate relatively simple localization topology. Consideration 3 is related to the computational efficiency. As the smooth solution away from the localization can be efficiently obtained by the RK approximation, the domain of influence of the NN approximation is controlled to be in the regions near localizations. The NN approximation will be designed such that the domain of influence is controlled by adjustable NN parameters that are automatically determined by the optimization procedures. For example, when multiple sparsely-distributed localization clusters are populated in the domain, as shown in Figure 3, the domain of influence in the NN approximation is sparsely distributed near the localization clusters. In the construction of the NN approximation space that satisfies Considerations 1-3, parameters that control the location, orientation, and shape of a localization are included, and these parameters are determined by the potential energy minimization. In the next subsection, a design of a block-level NN approximation satisfying Considerations 1-3 is proposed.

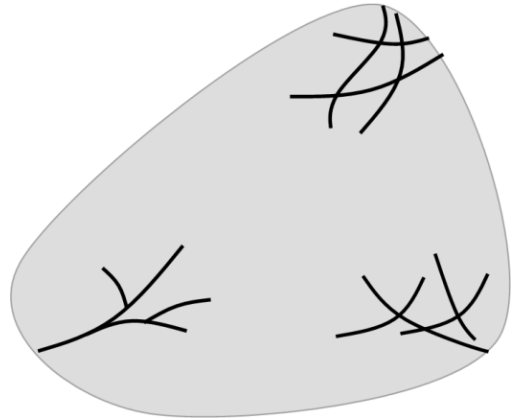

Figure 3. A domain with multiple localization regions

## 3.2. Block-level approximation

The proposed NN approximation is constructed by the block-level NN approximations as follows.

$$u^{NN}(\mathbf{x}) = \sum_{B=1}^{N_B} b_B^{NN}(\mathbf{x}; \mathbf{W}), \tag{23}$$

where $b_B^{NN}$ is $B$-th block-level approximation with unknown weights and biases lumped into $\mathbf{W}$, and $N_B$ is the number of blocks. Each NN block is constructed to capture a certain portion of the localizations distributed throughout the domain and to enrich the solution near localizations. Here, the introduction of NN blocks is intended to reduce the number of unknown weights and biases in representing a localization network. Also, the superposition of multiple block-level approximations is performed to capture complex localization topology.

In this work, the block-level approximation takes the following form:

$$b_B^{NN}(\mathbf{x}; \mathbf{W}) = \sum_{K=1}^{N_K} \hat{\phi}_{KB}(\mathbf{x}; \mathbf{W}^L, \mathbf{W}^S) p(\mathbf{x}; \mathbf{W}_{KB}^P) \tag{24}$$

where $p$ is a monomial function and $\hat{\phi}_{KB}$ is the $K$-th normalized NN kernel function of the $B$-th NN block, both are to be constructed automatically with the neural networks. The role of the kernel function $\hat{\phi}_{KB}$ is to capture the location, orientation, and the transition shape of the localization. Here, $\hat{\phi}_{KB}$ involves two sets of unknown parameters: the localization location and orientation parameter set, $\mathbf{W}^L \equiv \{\mathbf{W}_B^L\}_{B=1}^{N_B}$, and the kernel shape parameter set, $\mathbf{W}^S \equiv \{\{\mathbf{W}_{KB}^S\}_{K=1}^{N_K}\}_{B=1}^{N_B}$. Meanwhile, the monomial function, $p(\mathbf{x}; \mathbf{W}_{KB}^P) = \bar{\mathbf{H}}^T(\mathbf{x})\mathbf{W}_{KB}^P$, determines the order of completeness in the neural network function space with a set of monomial basis functions, $\bar{\mathbf{H}}(\mathbf{x}) = [1, x_1, x_2, \cdots, x_1^p x_2^q x_3^r, \cdots, x_3^n]^T$, and their corresponding weights, $\mathbf{W}_{KB}^P = [\bar{w}_{000}^{KB}, \bar{w}_{100}^{KB}, \bar{w}_{010}^{KB}, \bar{w}_{pqr}^{KB}, \cdots \bar{w}_{00N}^{KB}]^T$ with $p + q + r \leq n$ for 3-dimension.

The other NN parameter sets, $\mathbf{W}_B^L$ and $\mathbf{W}_{KB}^S$, for enhancement of the RK approximation for capturing localizations will be introduced next.

### 3.2.1. NN kernel functions by $\mathbf{W}_{KB}^S$

Let us consider the following one-dimensional kernel function for the NN approximation that is constructed by the multiplication of two regularized step functions $\bar{\phi}_1$ and $\bar{\phi}_2$ (see Figure 4): for $K$-th NN kernel function of block $B$,

$$\phi(y; \mathbf{W}_{KB}^S) = \bar{\phi}_1(y; \{\bar{y}_1^{KB}, c_1^{KB}, \beta_1^{KB}\})\bar{\phi}_2(y; \{\bar{y}_2^{KB}, c_2^{KB}, \beta_2^{KB}\}). \tag{25}$$

where $y$ is the 1-D coordinate and $\mathbf{W}_{KB}^S = \{\bar{y}_i^{KB}, c_i^{KB}, \beta_i^{KB}\}_{i=1}^2$ is a set of shape control parameters to be determined by the NN optimization. For the remaining section, the subscript $i$ will denote that the corresponding quantity is associated with $i$-th regularized step function, $\bar{\phi}_i$.

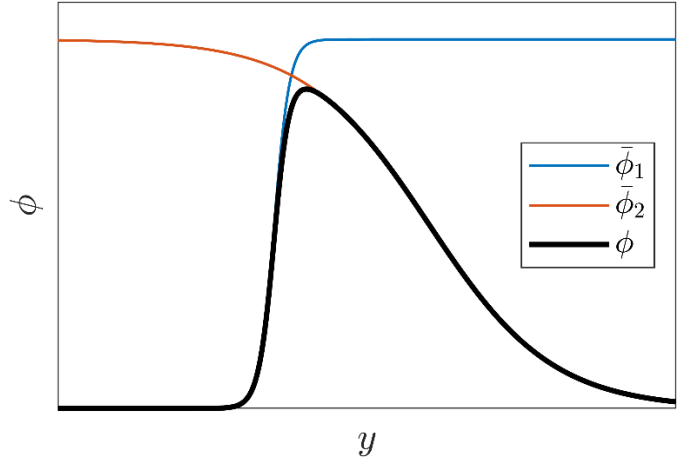

Figure 4. Schematic illustration of an NN kernel $\phi = \bar{\phi}_1 \bar{\phi}_2$: $\bar{\phi}_1$ and $\bar{\phi}_2$ produces transitions on the left and right sides of the kernel functions, respectively. See Figure 5 - Figure 9 for the effects of shape parameters.

In the following discussions, $\bar{\phi}_i$ takes the following form where the superscript $KB$ is dropped for brevity.

$$\bar{\phi}_i(y; \{\bar{y}_i, c_i, \beta_i\}) \equiv \bar{\phi}(z_i(y, \bar{y}_i, c_i); \beta_i) = S\left(z_i + \frac{1}{2}; \beta_i\right) - S\left(z_i - \frac{1}{2}; \beta_i\right),$$

$$z_i = (-1)^i (y - \bar{y}_i)/c_i, \quad i = 1, 2$$

(26)

where $z_i$ and $S(z; \beta)$ are the normalized coordinate and Softplus function, respectively. The Softplus function $S(z; \beta)$ is defined as

$$S(z; \beta) = \frac{1}{\beta} \log(1 + e^{\beta z}). \tag{27}$$

where $\beta$ controls the sharpness in the transition of derivative. Given values of $\beta$, $S(z; \beta)$ and $\bar{\phi}(z; \beta)$ are plotted in Figure 5 where larger $\beta$ values lead to sharper derivative transition.

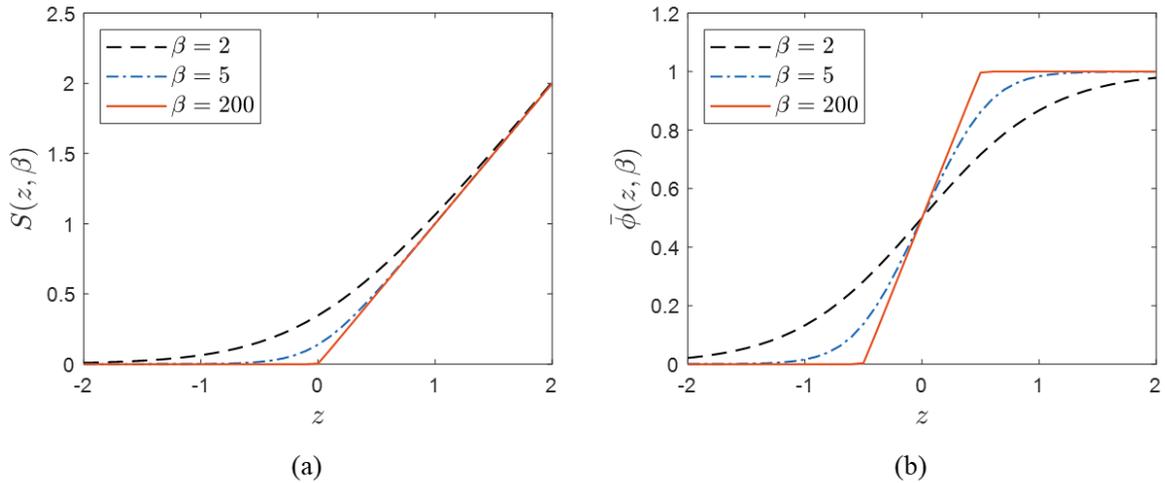

Figure 5. (a) $S(z,\beta)$ and (b) $\bar{\phi}(z,\beta)$

Figure 6 illustrates the aforementioned NN kernel function, as defined in Eq. (25), constructed with various $c_1$ and $c_2$ values while the other control parameters are fixed. Smaller $c_1$ and $c_2$ values yield more localized transitions in the kernel functions, as shown in Figure 6 (a) and (c); thus, more highly localized kernel function derivatives as shown in Figure 6 (b) and (d). Figure 7 displays kernels and their derivatives varied by $\beta_1$ while the other control parameters are fixed. As shown in Figure 7 (d), larger $\beta_1$ leads to sharper transition in the derivative; $\beta_1 = 200$ well approximates the strong discontinuities in the derivative, resulting in a kernel function that is suitable to achieve the Consideration 1 discussed in Section 3.1. Theoretically, with $\beta \to \infty$, the derivative possesses strong discontinuities. The NN parameter $\bar{y}_i$ controls the position of $\bar{\phi}_i$, and the interval, $\bar{y}_2 - \bar{y}_1$, affects the localization width of $\phi$, as shown in Figure 8. Note that both $c_i$ and $\bar{y}_i$ affect the domain of influence but in different ways: $\bar{y}_i$ changes the localization width by varying the distance between the left and right transition zones without altering the sharpness of the transition while $c_i$ affects the localization intensity by varying the sharpness of the transition as shown in Figure 6. In this work, this kernel is called the *neural network (NN) kernel*.

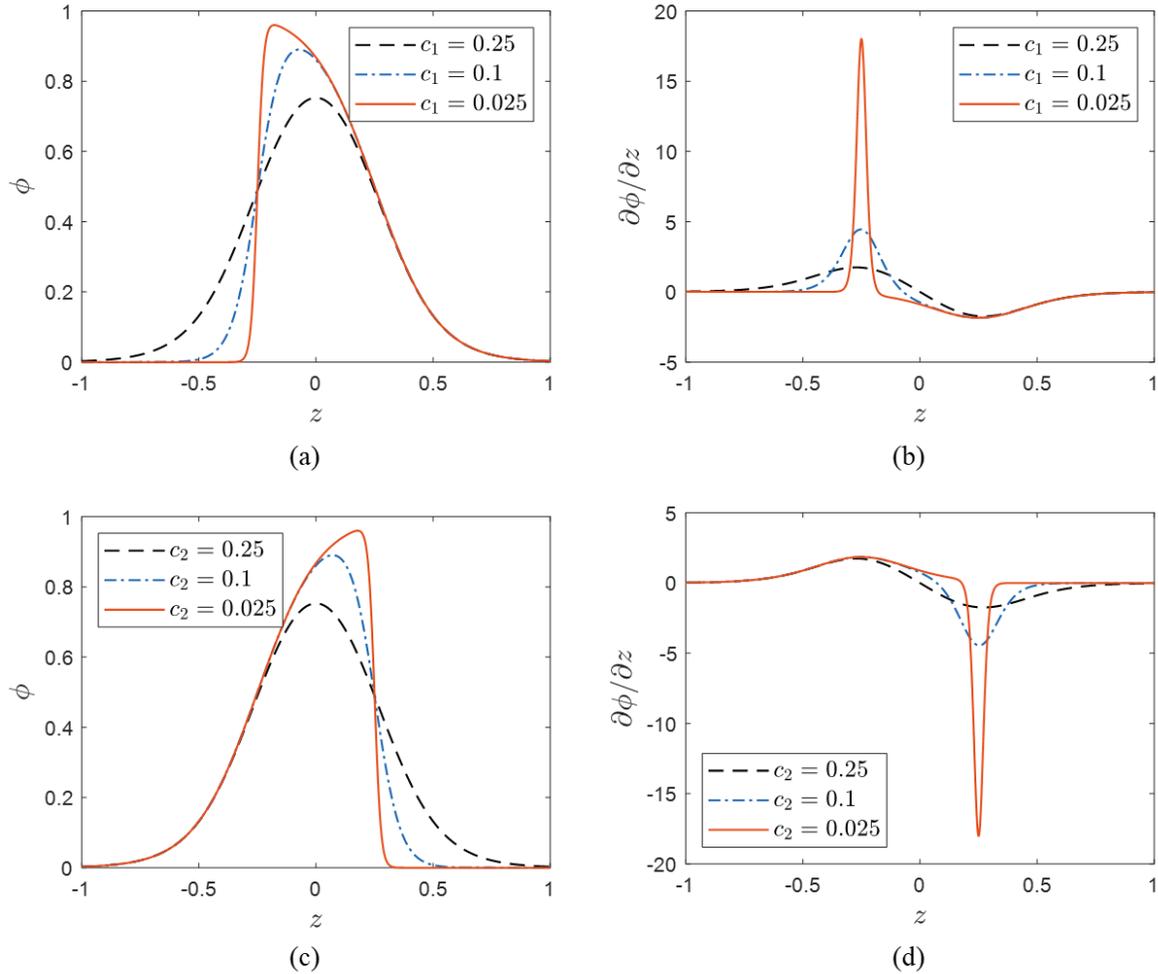

Figure 6. Kernel functions constructed with various $c_i$, given $\{\bar{y}_1, \bar{y}_2, \beta_1, \beta_2\} = \{-0.25, 0.25, 2.0, 2.0\}$: (a) kernel functions with various $c_1$ with $c_2 = 0.25$, (b) the derivatives of the kernel functions shown in (a), (c) kernel functions with various $c_2$ with $c_1 = 0.25$, and (d) the derivatives of the kernel functions shown in (c)

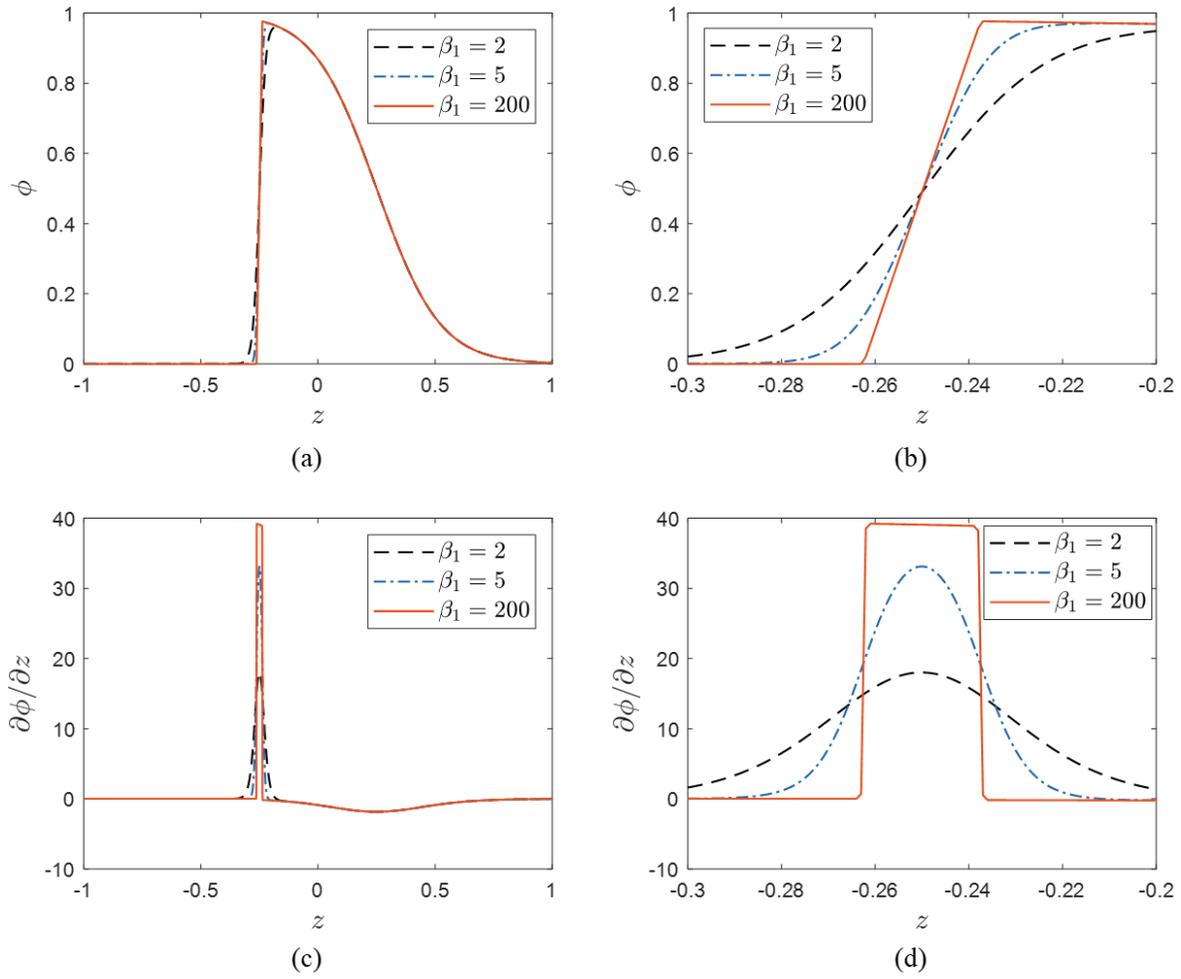

Figure 7. Kernel functions and their derivatives constructed with various $\beta_1$, given $\{\bar{y}_1, \bar{y}_2, c_1, c_2, \beta_2\} = \{-0.25, 0.25, 0.025, 0.25, 2.0\}$: (a) kernel functions, (b) kernel functions zoomed around $z = 0.25$, (c) kernel function derivatives, and (d) kernel function derivatives zoomed-in around $z = -0.25$

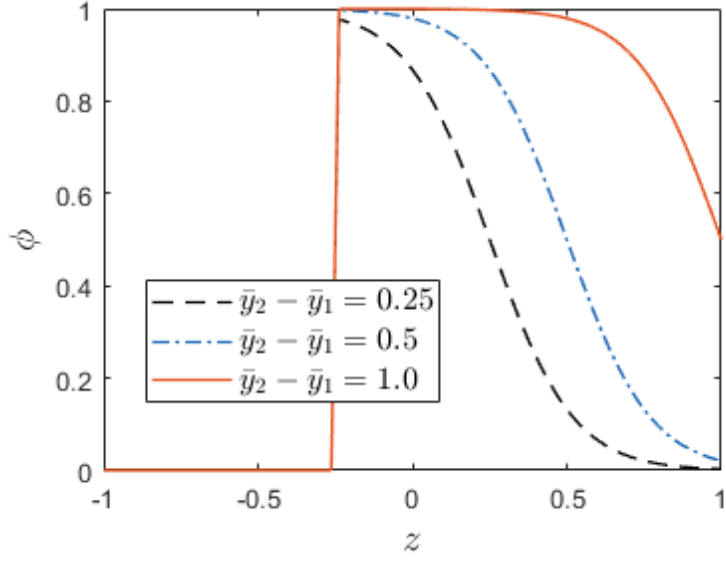

Figure 8. Domain of influence controlled by $\bar{y}_2$

For a set of multiple kernel functions employed, $\{\{\phi(y;\mathbf{W}_{KB}^S)\}_{K=1}^{N_K}\}_{B=1}^{N_B}$ with $N_K$ denoting the number of NN kernel functions per NN block, the following normalization is considered to obtain the partition of unity property:

$$\hat{\phi}_{KB}(y) = \frac{\phi(y;\mathbf{W}_{KB}^S)}{\sum_{A=1}^{N_K}\sum_{I=1}^{N_B}\phi(y;\mathbf{W}_{IA}^S) + \epsilon} \qquad (28)$$

where $\hat{\phi}_{KB}(y)$ is the normalized NN kernel and $\epsilon$ is a small number introduced to prevent sparsely distributed NN kernel functions from having too large domains of influence (see Figure 9), so that the NN approximation only influences regions near localizations. $\epsilon$ must be small enough to approximate the partition of unity property for the region where the material point is covered by an enough number of NN kernel functions. In this work, $\epsilon = 10^{-3}$ is used.

*Remarks 3.1:*

1. Note that $\epsilon$ is not a trainable parameter; the value of $\epsilon$ does not evolve during the potential energy minimization.

2. The localization width and intensity, and the transition shape of the NN kernel function are controlled by the aforementioned shape control parameter $\mathbf{W}_{KB}^S$ that evolves in the potential energy minimization process: the localization width is controlled by $\bar{y}_2 - \bar{y}_1$ as shown in Figure 8, the localization intensity is controlled by $c_i$ as shown in Figure 6, and the localization transition shape is controlled by $\beta_i$ as shown in Figure 7.

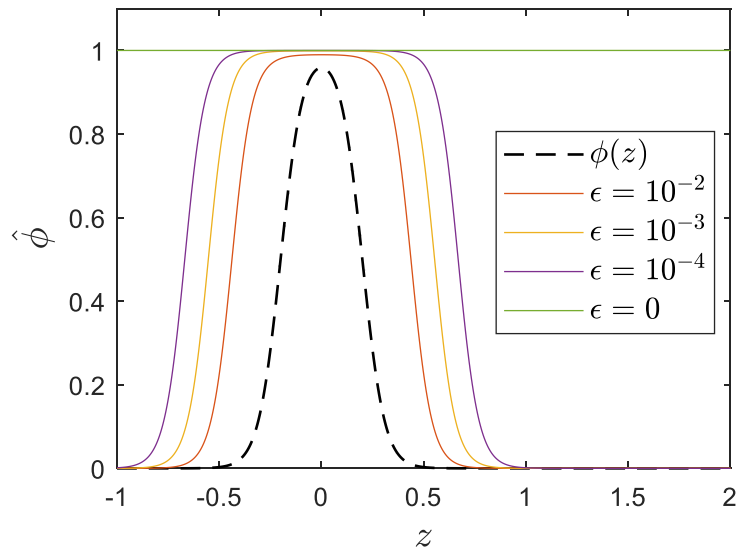

Figure 9. Effect of $\epsilon$ in kernel normalization in case that $N_K = 1$. When $\epsilon = 0$, $\hat{\phi}$ is always unity and has an infinite domain of influence.

The extension of the NN kernel to two-dimension with coordinate $\mathbf{y} = (y_1, y_2)$ is straightforward: for the $K$-th kernel in the $B$-th block,

$$\phi(\mathbf{y}; \mathbf{W}_{KB}^S) = \prod_{\alpha=1}^{2} \prod_{i=1}^{2} \bar{\phi}_i(y_\alpha; \mathbf{w}_{\alpha i}^{KB}) \tag{29}$$

where $\mathbf{w}_{\alpha i}^{KB} = \{\bar{y}_{\alpha i}^{KB}, c_{\alpha i}^{KB}, \beta_{\alpha i}^{KB}\}$ is the kernel shape parameter set associated for the $\alpha$-th direction and with $i$-th regularized step function. Similar to Eq. (26), the regularized step function, $\bar{\phi}_i$, takes the following form where the superscript, $KB$, is dropped for brevity:

$$\bar{\phi}_i(y_\alpha; \mathbf{w}_{\alpha i}) \equiv \bar{\phi}(z_{\alpha i}(\bar{y}_{\alpha i}, c_{\alpha i}); \beta_{\alpha i}), \ \alpha = 1, 2, i = 1, 2 \tag{30}$$

where $\bar{\phi}$ is defined in Eq. (26), and $z_{\alpha i} = (-1)^i (y_\alpha - \bar{y}_{\alpha i})/c_{\alpha i}$ is similarly introduced.

### 3.2.2. Parametrization by $\mathbf{W}^L$

In multi-dimensional problems, highly complex localization topological patterns can be present. In this approach, the complicated localization patterns will be projected onto a parametric space in each NN block "A" with parametric coordinates, $\mathcal{P}^A: \mathbf{x} \to \mathbf{y}^A$, where $\mathbf{x} \in \mathbb{R}^D$ and $\mathbf{y}^A \in \mathbb{R}^D$ are physical coordinates and parametric coordinates of block A, respectively. Figure 10 provides a schematic illustration showing a localization pattern (red curves) in the physical coordinate and in the parametric coordinate. For complicated localization topological patterns, a superposition of multiple block-level NN approximations, each with a mapping $\mathcal{P}^A: \mathbf{x} \to \mathbf{y}^A$, is introduced.

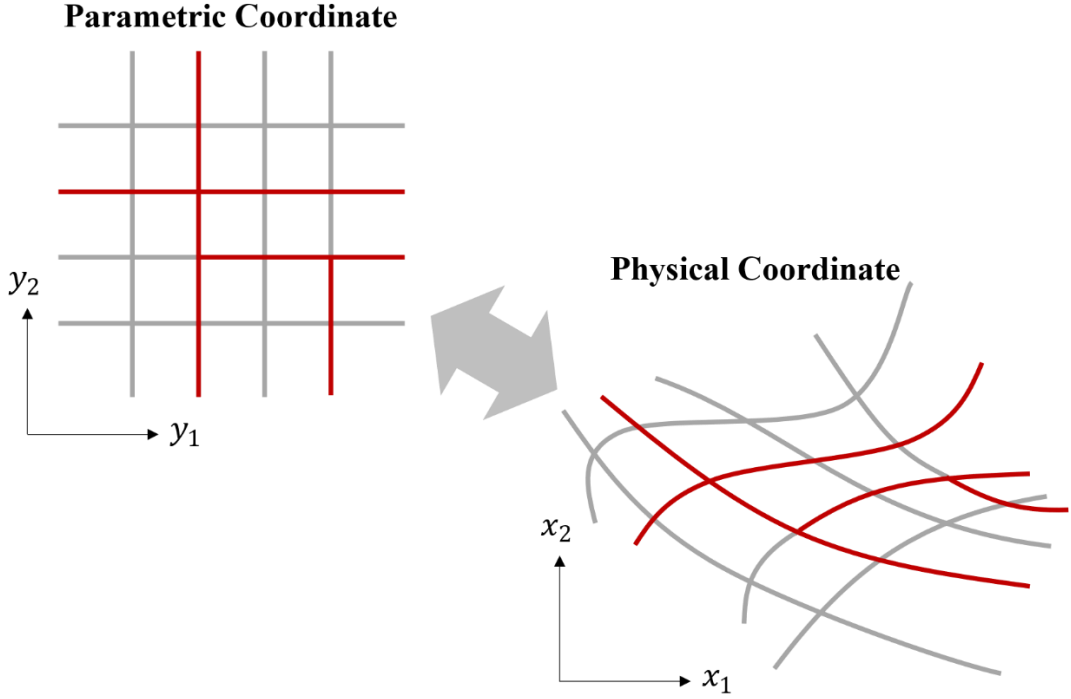

Figure 10. Parametric and physical coordinates: the red curves in the physical coordinate denote discontinuities. The coordinates are parametrized as shown in the figure to the right.

For parametrization, multiple NN layers with hyperbolic tangent activation functions can be considered as follows.

$$\mathbf{y}(\mathbf{X}; \mathbf{W}_B^L) = \mathbf{h}_{n_L}\left(\cdot; \mathbf{W}_{Bn_L}^L\right) \circ \mathbf{h}_{n_L-1}\left(\cdot; \mathbf{W}_{Bn_L-1}^L\right) \circ \cdots \circ \mathbf{h}_1(\mathbf{x}; \mathbf{W}_{B1}^L) \qquad (31)$$

where $\mathbf{W}_B^L = \{\mathbf{W}_{B1}^L, \mathbf{W}_{B2}^L, \cdots, \mathbf{W}_{Bn_L}^L\}$ with $\mathbf{W}_{Bi}^L$ the weights and biases of layer $i$ of block $B$, $n_L$ is the number of layers employed in the parameterization network, and $\mathbf{h}_i$ of the $i$-th layer is defined as

$$\mathbf{h}_i(\boldsymbol{\xi}; \mathbf{W}_{Bi}^L) = \tanh\left(\mathbf{z}_i(\boldsymbol{\xi}; \mathbf{W}_{Bi}^L)\right), \qquad \text{for } i < n_L \tag{32}$$

$$\mathbf{h}_i(\boldsymbol{\xi}; \mathbf{W}_{Bi}^L) = \mathbf{z}_i(\boldsymbol{\xi}; \mathbf{W}_{Bi}^L), \qquad \text{for } i = n_L,$$

with

$$\mathbf{z}_i(\boldsymbol{\xi}; \mathbf{W}_{Bi}^L) = \boldsymbol{\Theta}^{Bi}\boldsymbol{\xi} + \boldsymbol{\beta}^{Bi} \tag{33}$$

where $\mathbf{W}_{Bi}^L = \{\boldsymbol{\Theta}^{Bi}, \boldsymbol{\beta}^{Bi}\}$ with the weight matrix $\boldsymbol{\Theta}^{Bi}$ and the bias vector $\boldsymbol{\beta}^{Bi}$.

*Remark.* Although it may be beneficial for the NN blocks to have their own parametric coordinates to capture complicated localization topology, a parametric coordinate can be shared by multiple NN blocks in general for computational efficiency. In this case, the NN kernel functions that belong to different NN blocks are defined in the same parametric coordinate.

### 3.2.3. Approximation ability of a single neural network block

As discussed in Section 3.2.2, the superposition of block-level NN parametrizations is introduced to capture the topological localization patterns. For two-dimensional problems, each NN block contains four NN kernels defined in (29)-(30) along with the parameterization network defined in (31)-(33). Figure 11 shows the topological geometries that can be captured by a single four-kernel NN block. The approximation $u^h$ is obtained by the standard $L_2$ minimization to obtain NN parameters $\mathbf{W} = \{\mathbf{W}^L, \mathbf{W}^S, \mathbf{W}^P\}$.

$$\min_{\mathbf{d},\mathbf{W}} \int_\Omega \left(u - u^h(\mathbf{d},\mathbf{W})\right)^2 d\Omega \qquad (34)$$

where $u$ is the target function and $u^h(\mathbf{d},\mathbf{W}) = u^{RK}(\mathbf{d}) + u^{NN}(\mathbf{W})$ is the approximation with the RK coefficient, $\mathbf{d}$, and the collection of unknown NN weights, $\mathbf{W}$. For the RK approximation, 100 uniformly distributed RK nodes are used. For the NN approximation, a single parametrization layer with 16 neurons and a linear NN basis is used. The total number of unknowns are 358 consisting of 200 RK coefficients and 158 unknown NN weights. The domain integration is performed by a 2×2 Gauss quadrature applied to 72×72 uniformly distributed integration cells.

As shown in Figure 11, the NN approximation, $u^{NN}$, with one four-kernel NN block successfully captures the sharp solution transition of very high gradient with an $L_2$ error of approximately $10^{-3}$ or below for three different topological geometries: 1) a curve without junctions, 2) 1 triple junction, and 3) 1 quadruple junction. The superposition of multiple block-level NN approximations is needed to capture higher order topological geometries such as 2 or more triple junctions or quadruple junctions. Discussions will be made in the next section for the convergence performance of NN-RK approximation with respect to the number of neurons and NN blocks.

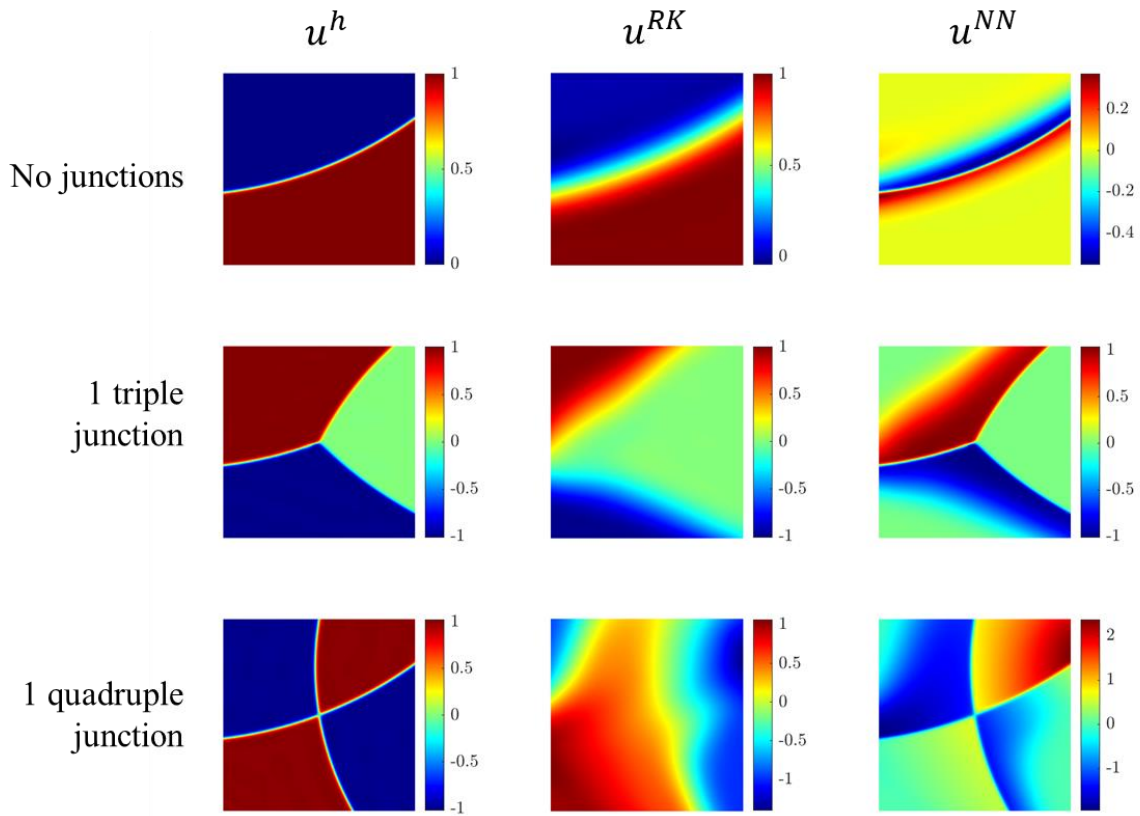

Figure 11. Types of topological geometries captured by a single block-level NN approximation with four NN kernels

### 3.2.4. Convergence performance

To study convergence of the proposed method, three target functions (TF1, TF2, and TF3) are constructed for a domain $\Omega = [-1,1] \times [-1,1]$, as shown in Figure 12, with target function expressions given in Appendix A. TF1 and TF2 contain a narrow transition region that follows a circular arc. These functions are constructed such that the transition width normal to the curve is constant, $w = 0.04$. Each side of the transition band takes on distinct function values. For TF1, the constant values on either side of the transition zone are set as 0 and 1 below and above the transition zone respectively. For TF2, the function is 0 below the transition zone and varies with respect to x above the transition zone as $\cos(\pi(x + 1)/4)$, inducing a varied transition zone jump across the domain. TF3 decomposes the domain into five constant-valued sections separated by transition zones, which are defined by intersecting circular arcs and the domain boundaries. A total of three triple-junctions are formed using seven arc transition zones, each of constant width $w = 0.04$. The exact equations used to construct TF3 can be found in Appendix A.

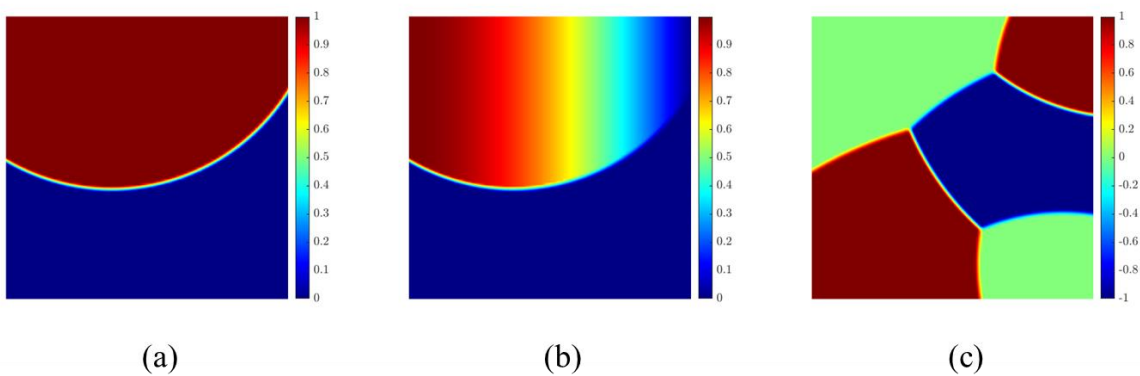

(a) (b) (c)

Figure 12. Target functions constructed for convergence studies: (a) constant values on either side of the transition zone, (b) constant value on one side of the transition zone and varying with respect to x-direction on the other, (c) three triple-junctions with five constant-valued zones

Three convergence studies are explored with varied number of neurons ($N_{NR}$) in the parametrization layer and number of blocks ($N_B$). The number of kernel functions ($N_K$) in each NN block is fixed to be $N_K$=4. For TF1 and TF2, due to the simplicity of the localization topology, only $N_{NR}$ is varied while keeping $N_B$=1 and $N_K$=4 in the convergence study. For TF3, due to the complexity of localization topological pattern, increasing the number of NN blocks is more effective in the NN-based approximation, and hence $N_B$ is varied in the convergence study, while keeping $N_{NR}=32$ and $N_K=4$. Table 1 summarizes the settings of each study.

Table 1. Convergence of the NN based target function approximation

|        | TF1 | TF2 | TF3 |
|--------|-----|-----|-----|
| Case 1 | $N_{NR}=8$, $N_K=4$, $N_B=1$ | $N_{NR}=8$, $N_K=4$, $N_B=1$ | $N_{NR}=32$, $N_K=4$, $N_B=1$ |
| Case 2 | $N_{NR}=16$, $N_K=4$, $N_B=1$ | $N_{NR}=16$, $N_K=4$, $N_B=1$ | $N_{NR}=32$, $N_K=4$, $N_B=2$ |
| Case 3 | $N_{NR}=32$, $N_K=4$, $N_B=1$ | $N_{NR}=32$, $N_K=4$, $N_B=1$ | $N_{NR}=32$, $N_K=4$, $N_B=4$ |
| Case 4 |     | $N_{NR}=64$, $N_K=4$, $N_B=1$ | $N_{NR}=32$, $N_K=4$, $N_B=8$ |
| Case 5 |     | $N_{NR}=128$, $N_K=4$, $N_B=1$ |     |

The domain is discretized using 10x10 uniformly spaced RK nodes and 72x72 Gauss cells with 2x2 Gauss integration for domain integration in (34). For all convergence studies discussed in this section, the $L_2$ error norm is integrated using 8x8 Gauss integration and the same 72x72 Gauss cells. The convergence rates are calculated with respect to the square root of the varied NN variables.

For the approximation of TF1, $N_{NR}$ is varied as 8, 16, and 32. The observed $L_2$ error norm decreases as the number of neurons is increased at an average rate of convergence of 2.282, as seen in Figure 13.

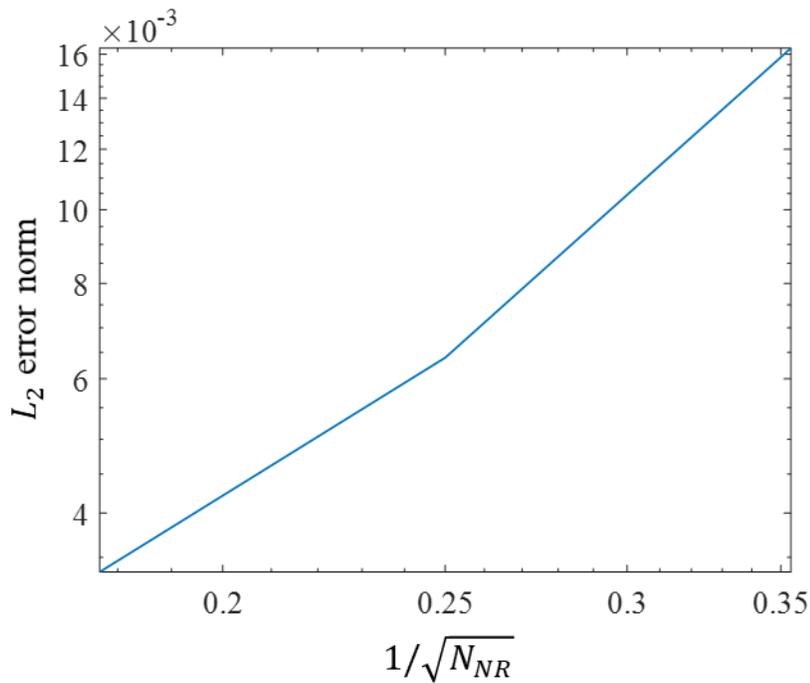

Figure 13. Convergence plot for TF1: average convergence rate of 2.282

For the approximation of TF2, and $N_{NR}$ is varied as 8, 16, 32, 64, and 128. The average convergence rate for the first four cases is 0.430, but the error reduces significantly as $N_{NR}$ is increased to 128, leading to an average convergence rate of 1.620 with all 5 cases considered. TF2 presents a more challenging function to approximate than TF1, as the neural network must learn the transition zone location as well as the varying magnitude of the jump. As such, a more refined neural network is needed in the parametrization layer. Once a sufficiently large number of neurons is incorporated, the error reduces significantly.

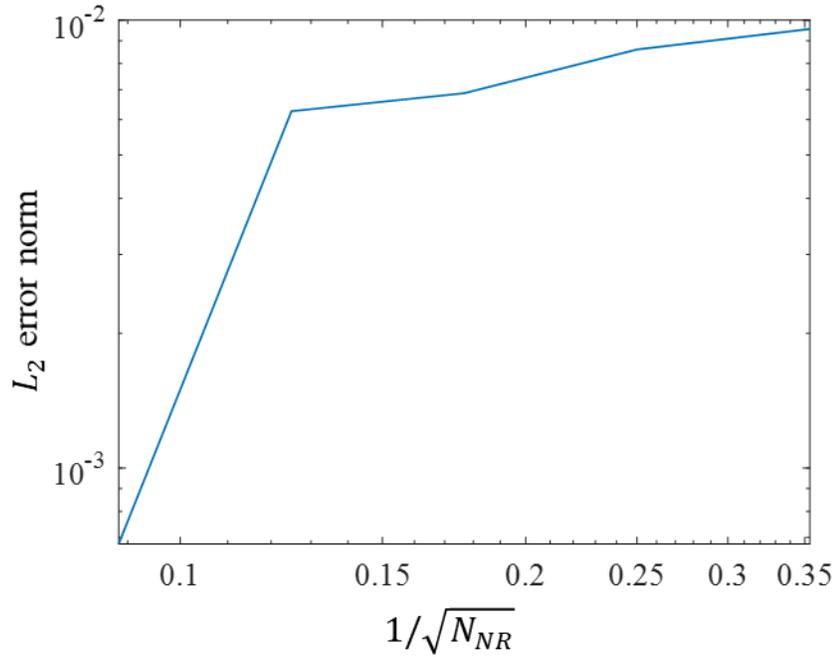

Figure 14. Convergence plot for Study 1-2: average convergence rate of 1.620

In the approximation of TF3, $N_B$ is varied across 1, 2, 4, and 8. The average rate of convergence is shown in Figure 15 as 3.578. The $L_2$ error norm monotonically decreases as $N_B$ is increased, as shown in Figure 15. The $L_2$ error decreases most significantly between the 2- and 4-block cases. Due to the existence of three triple-junctions in TF3, increasing $N_B$ beyond 4 appears to be less effective.

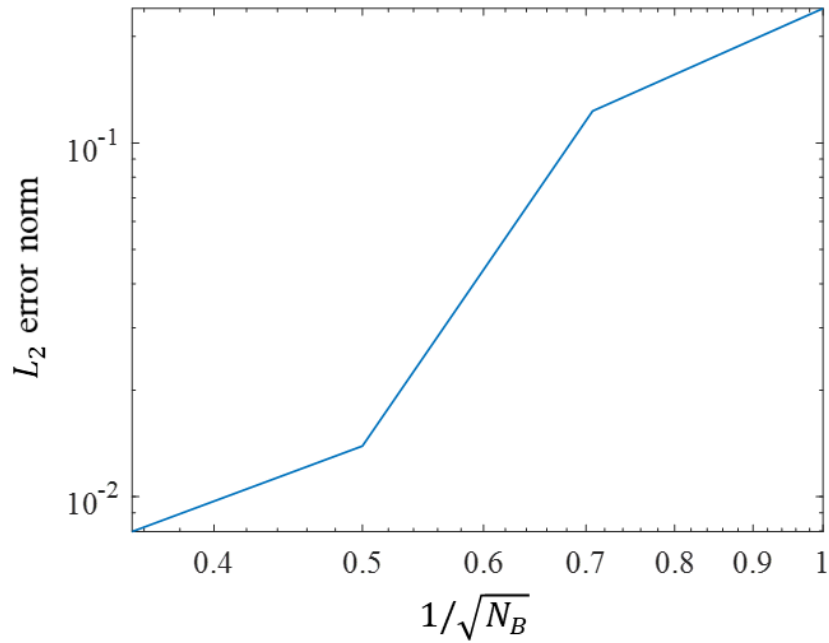

Figure 15. Convergence plot for Study 1-3: average convergence rate of 3.578. Convergence rates of 1.925 and 1.607 between 1- and 2-block cases and 4- and 8-block cases respectively.

As demonstrated in Figure 16, although the location of the transition zones can be generally captured with just one block, the narrow transition zones are not well captured until 4 and 8 NN blocks are used. Once a sufficient number of blocks is employed, the convergence slows down. Introducing more NN blocks into the neural network architecture increases the number of independent parametric coordinates to capture the complex localization topological patterns.

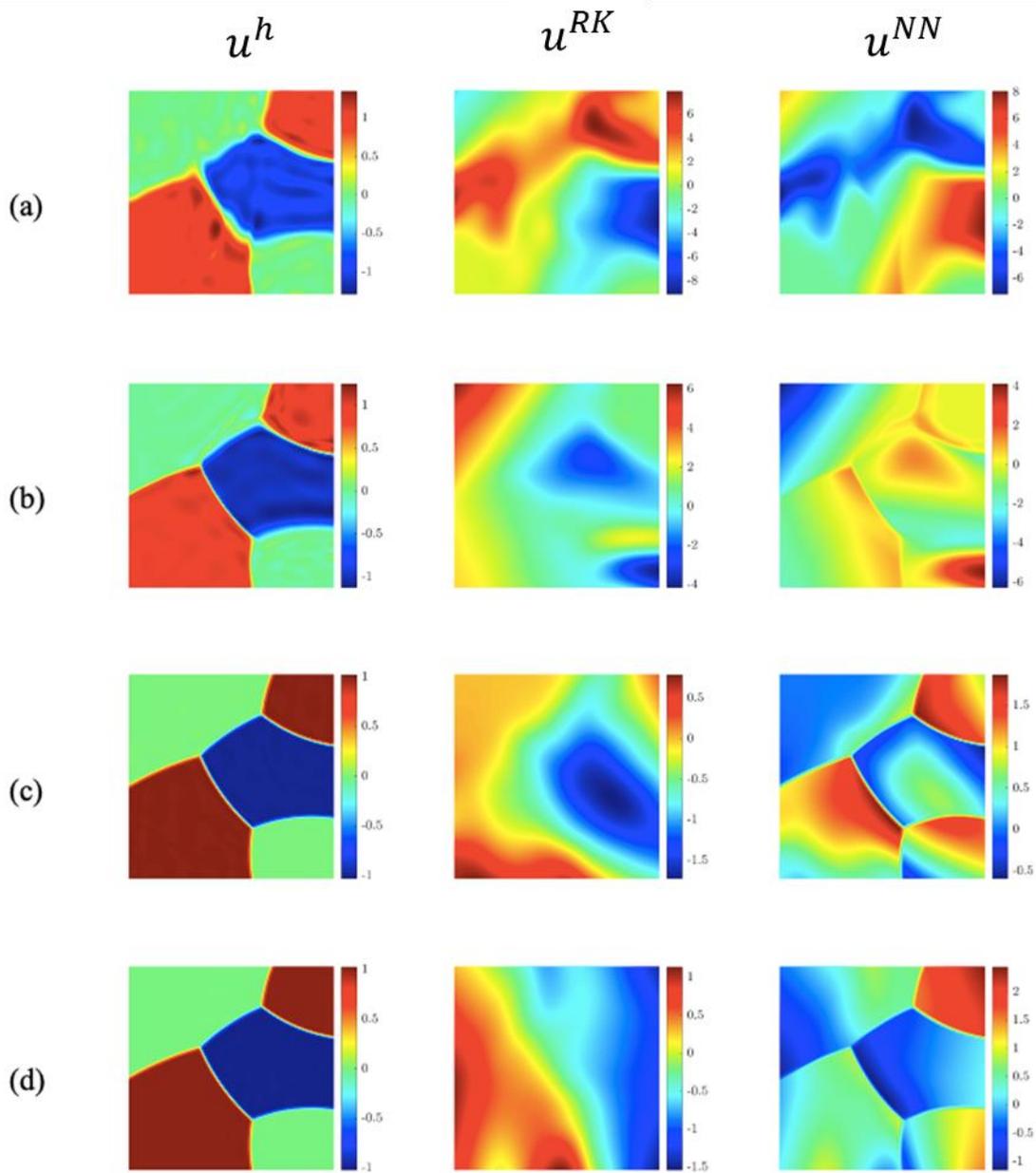

Figure 16. Converged numerical solution of TF3 with various $N_B$: (a) 1, (b) 2, (c) 4, (d) 8

### 3.3. Regularization

To ensure the proposed NN-RK approximation is discretization-insensitive when applied to damage mechanics, a regularization is introduced into the NN kernel function. Although the NN kernel contains the sharpness parameter $c$ in Eqs. (26) and (30), the localization width is also affected by the parametrization, more specifically, the gradient of the parametric coordinate $\mathbf{y}$ with respect to the physical coordinate $\mathbf{x}$. In this subsection, the parameters affecting the sharpness of kernel function in the parameterization are to be properly scaled by modifying Eqs. (26) and (30), so that the localization width is solely determined by the parameter $c$. This introduces a length scale to the physical parameters that can then be used to define an objective energy dissipation.

To analyze the influence of the parametrization on the transition sharpness, consider the first order Taylor expansion of $y_i(\mathbf{x})$ as

$$y_i(\mathbf{x}) \approx y_i(\bar{\mathbf{x}}) + (\mathbf{x} - \bar{\mathbf{x}}) \cdot \left.\frac{\partial y_i}{\partial \mathbf{x}}\right|_{\mathbf{x}=\bar{\mathbf{x}}} \tag{35}$$

Let us define $\widehat{H}$ as

$$\widehat{H} \equiv 1/\left\|\frac{\partial y_i}{\partial \mathbf{x}}\right\|_{\mathbf{x}=\bar{\mathbf{x}}} \tag{36}$$

where $\|\cdot\|$ is the Euclidean norm. Introducing (35) into the normalized coordinate $z$ in the regularized step function $\bar{\phi}$ in (26) and (30),

$$\frac{y_i - \bar{y}_i}{c} \approx \frac{(\mathbf{x} - \bar{\mathbf{x}}) \cdot \left(\frac{\partial y_i}{\partial \mathbf{x}}\bigg|_{\mathbf{x}=\bar{\mathbf{x}}}\right)}{c} = \frac{(\mathbf{x} - \bar{\mathbf{x}}) \cdot \left(\widehat{H}\frac{\partial y_i}{\partial \mathbf{x}}\bigg|_{\mathbf{x}=\bar{\mathbf{x}}}\right)}{c\widehat{H}}, \tag{37}$$

where $\bar{y}_i = y_i(\bar{\mathbf{x}})$. Note that $\widehat{H}\frac{\partial y_i}{\partial \mathbf{x}}\big|_{\mathbf{x}=\bar{\mathbf{x}}}$ in (37) is a unit vector, and $(\mathbf{x} - \bar{\mathbf{x}}) \cdot \left(\widehat{H}\frac{\partial y_i}{\partial \mathbf{x}}\big|_{\mathbf{x}=\bar{\mathbf{x}}}\right)$ is a projection of $(\mathbf{x} - \bar{\mathbf{x}})$ onto a unit vector in the direction $\frac{\partial y_i}{\partial \mathbf{x}}\big|_{\mathbf{x}=\bar{\mathbf{x}}}$. Therefore, $(\mathbf{x} - \bar{\mathbf{x}}) \cdot \left(\widehat{H}\frac{\partial y_i}{\partial \mathbf{x}}\big|_{\mathbf{x}=\bar{\mathbf{x}}}\right)$ is independent of the scaling between the physical coordinate and parametric coordinate. For the sharpness of the NN kernel functions in the physical coordinate to be controlled solely by the sharpness parameter $c$ and be independent of the mapping between the physical and parametric coordinates, the NN kernel functions in Eq. (26) and (30) are modified as follows.

$$\bar{\phi}(z;\beta) = S\left(z + \frac{1}{2};\beta\right) - S\left(z - \frac{1}{2};\beta\right), \qquad z = \frac{(y - \bar{y})\widehat{H}}{c}, \tag{38}$$

$$\bar{\phi}(z_{\alpha i};\beta_{\alpha i}) = S\left(z_{\alpha i} + \frac{1}{2};\beta_{\alpha i}\right) - S\left(z_{\alpha i} - \frac{1}{2};\beta_{\alpha i}\right),$$
$$z_{\alpha i} = \frac{(-1)^i(y_\alpha - \bar{y}_{\alpha i})\widehat{H}}{c_{\alpha i}}, \qquad \alpha = 1,2, \qquad i = 1,2 \tag{39}$$

Here, the sharpness in the NN kernel function $\bar{\phi}$ is entirely controlled by $c$. With this modification, imposing a numerical length scale $\ell$ as a lower bound of $c$ regularizes the solution in problems with strain localization.

To obtain an objective energy dissipation, the relation among the physical parameters $\kappa_c$

introduced in (6), the fracture energy $G_F$, and the tensile strength $f_t$, and the numerical length scale $\ell = c$ is determined following Wei and Chen (2018)[28]:

$$\kappa_c = \frac{2G_F}{f_t \ell} = \frac{2G_F}{E\kappa_0 \ell} \qquad (40)$$

# 4. Numerical implementation

## 4.1. Network structures

Figure 17 describes the neural network structures considered in this work. Given $n_L$, $N_{NR}$, $N_K$, $N_B$, and $N_P$ (denoting the number of layers of the parametrization network, the number of neurons per layer of the parametrization network, the number of NN kernels per block, the number of NN block, and the number of RK nodes, respectively), the entire network provides the map $\mathbb{R}^d \to \mathbb{R}$ with input $\mathbf{x} \in \mathbb{R}^d$ and output $u^h \in \mathbb{R}$. As the RK approximation is used in approximating the smooth part of the solution, a relatively coarse, predetermined discretization is sufficient. Since in this approach the RK shape functions are designed not to evolve throughout the simulation, the shape functions $\{\Psi_I(\mathbf{x})\}_{I=1}^{NP}$ are precomputed and directly entered into the network as the input as shown in Figure 17. The RK network has the RK generalized coordinates as its weights. On the other hand, the NN approximation directly takes the coordinate $\mathbf{x}$ as its input. Instead of employing a densely connected deep neural network, multiple network blocks are constructed in parallel, as discussed in Section 0, which introduces sparsity to the system of equations associated with the weights obtained by an optimization solver.

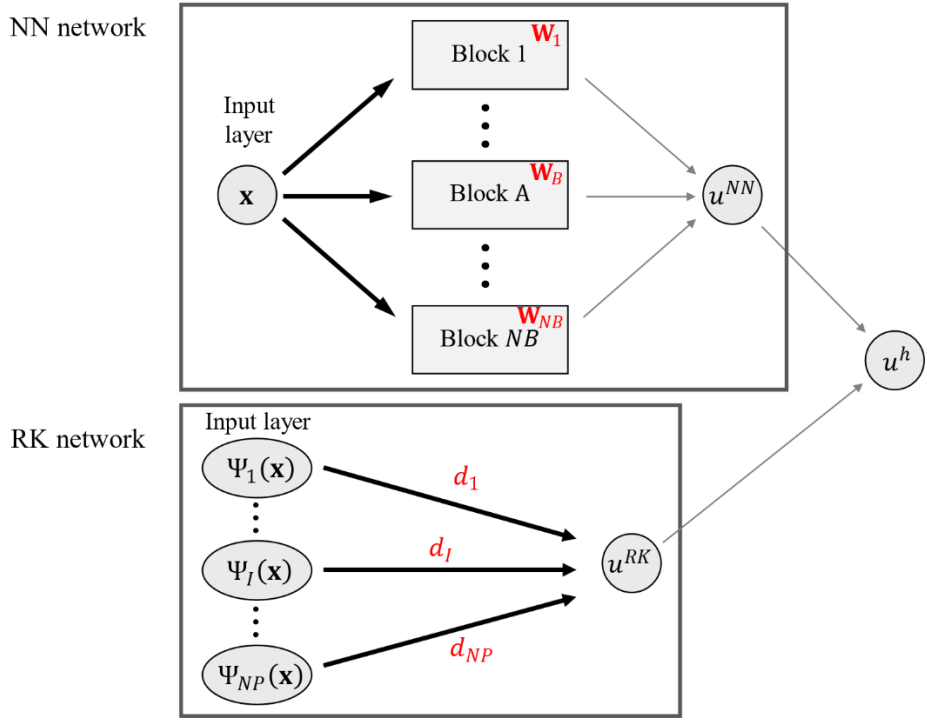

Figure 17. Entire network structure that incorporates the RK approximation and the NN approximation. The unknown parameters associated with each part of the network are shown in red. The thick black arrows denote that the network connections entail unknown parameters and the thin grey arrows denote that the weight of the associated connection is one (1) and do not change throughout the computation.

The network structure of a block $B$ is described in Figure 18. All the unknown parameters $\mathbf{W}^L$, $\mathbf{W}^S$, and $\mathbf{W}^P$ introduced in Section 0 are involved in each sub-block and solved by the minimization procedure. Here the parameterization network involves the localization location and orientation parameters set $\mathbf{W}^L$, with it outputs passing into the sub-block for NN kernel to construct the localization shape parameter set $\mathbf{W}^S$, while the polynomial basis set $\mathbf{W}^P$ sub-block is independently constructed. The details of the NN kernel sub-block are given in Figure 19 where the localization shape parameters $\mathbf{W}^S$ is included. In order to regularize the NN approximation as discussed in Section 3.3, the built-in function of Tensorflow

*tf.keras.constraints.MinMaxNorm* is utilized to impose the lower bound of $c_{\alpha i}$. The NN operation details of polynomial sub-block to obtain the polynomial basis functions $\mathbf{W}^P$ are presented in Figure 20.

The neuron-wise multiplication of the polynomial layer and kernel function layer is performed by employing *tensorflow.keras.layers.multiply* available in TensorFlow. Figure 21 shows how the $u^h$ derivatives are computed for use in solving the potential energy minimization problem. To compute the spatial derivatives of $u^{NN}$, the automatic differentiation function provided by TensorFlow is utilized. To compute the spatial derivatives of $u^{RK}$, the input of the RK network, $\Psi_I(\mathbf{x})$ in Figure 21 is replaced by the pre-computed shape function derivatives $\Psi_{I,i}(\mathbf{x})$.

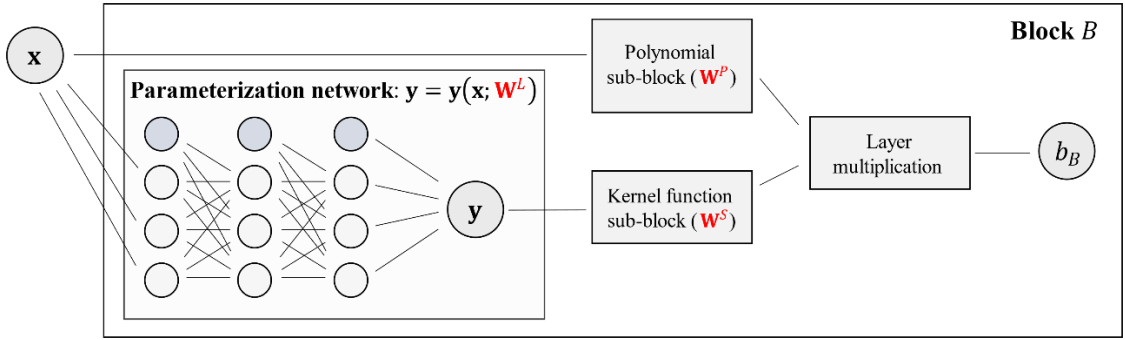

Figure 18. Network structure for block $B$. The unknown parameters introduced in each part are denoted in red color.

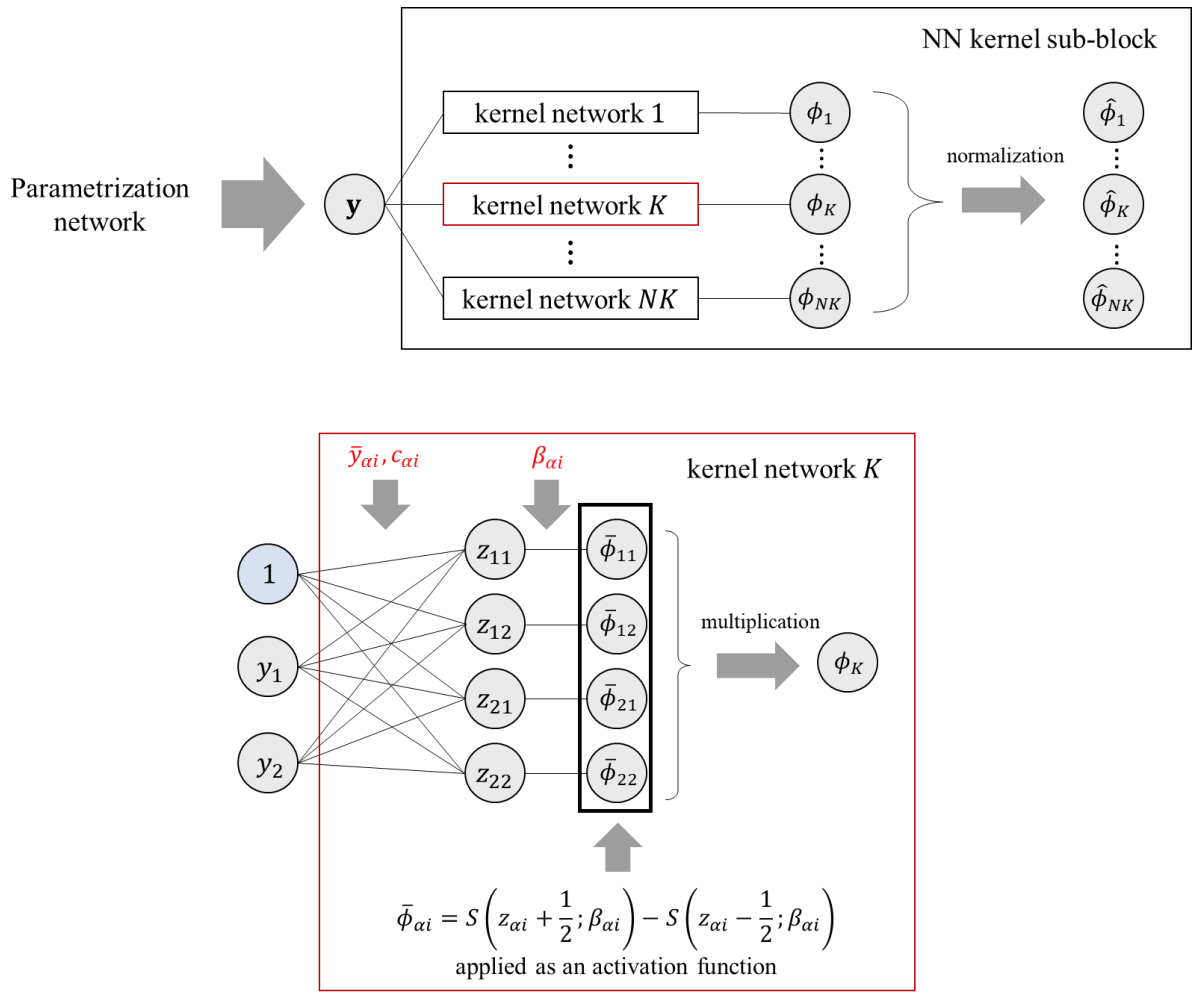

Figure 19. NN kernel sub-block: the shape control parameters $\mathbf{W}^S_{\alpha i} = \{\bar{y}_{\alpha i}, c_{\alpha i}, \beta_{\alpha i}\}$ are written in red.

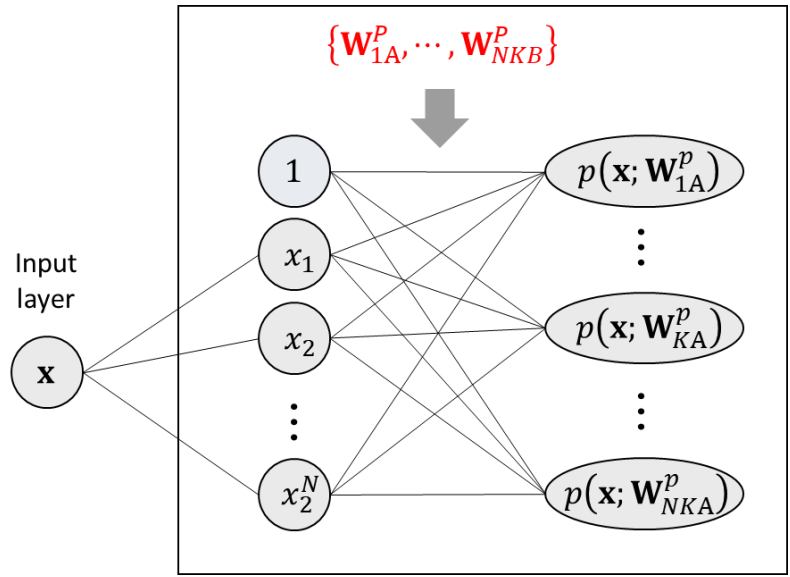

Figure 20. Polynomial sub-block details: The thick black lines denote that the network connections entail unknown parameters and the thin grey arrows denote that the weight of the associated connection is one (1) and do not change throughout the computation.

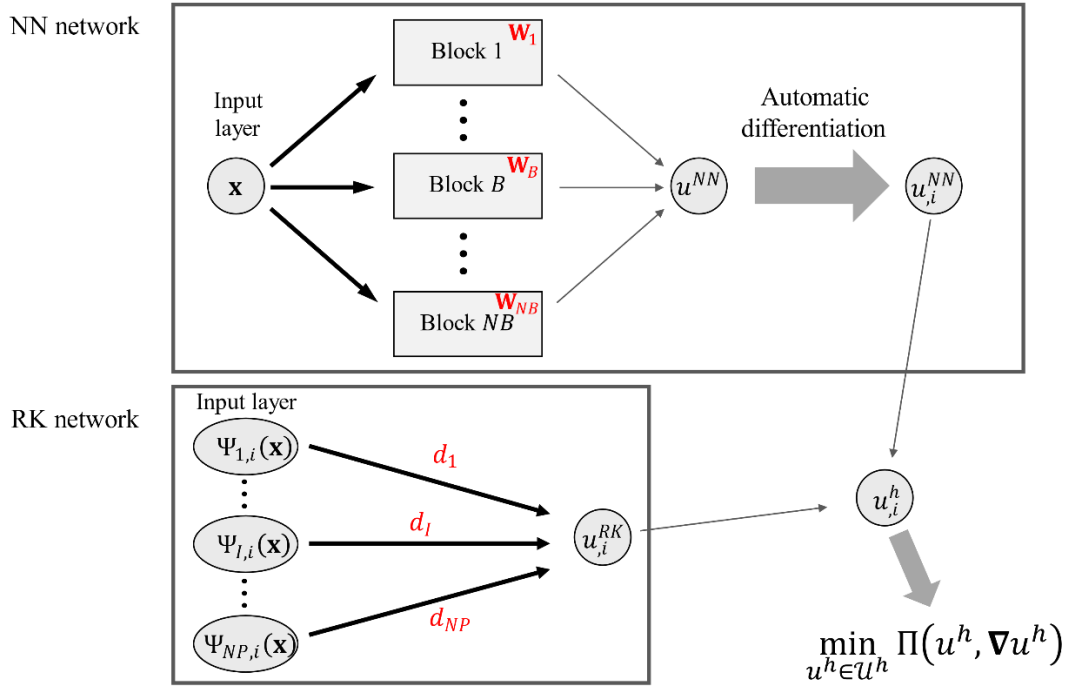

Figure 21. Computation of solution derivative $u_{,i}^h$: The thick black lines denote that the network connections entail unknown parameters and the thin grey arrows denote that the weight of the associated connection is one (1) and do not change throughout the computation.

### 4.2. Solution procedures

The optimization problem defined in (1) is employed to obtain the NN-RK approximated solution, $\mathbf{U}^h(\mathbf{d}, \mathbf{W})$, as follows:

$$\min_{\mathbf{d},\mathbf{W}} \Pi\left(\mathbf{U}^h(\mathbf{d}, \mathbf{W})\right) = \int_\Omega \psi\left(\mathbf{U}^h(\mathbf{d}, \mathbf{W})\right) d\Omega - F\left(\mathbf{U}^h(\mathbf{d}, \mathbf{W})\right), \tag{41}$$

where $\mathbf{U}^h(\mathbf{d}, \mathbf{W}) = \mathbf{U}^{RK}(\mathbf{d}) + \mathbf{U}^{NN}(\mathbf{W})$ is the collection of approximations with the RK coefficient set, $\mathbf{d}$, and the neural network weight set, $\mathbf{W} = \{\mathbf{W}^L, \mathbf{W}^S, \mathbf{W}^P\}$, and $\psi$ and $F$ are strain energy density and external work, respectively. The potential energy minimization problem is solved in two stages.

A. <u>RK Solution Stage</u>

In the first stage, the problem is solved only for the RK coefficients to find the initial guess of the RK coefficients $\bar{\mathbf{d}}$ as follows:

$$\bar{\mathbf{d}} = \underset{\mathbf{d}}{\operatorname{argmin}} \Pi\left(\mathbf{U}^{RK}(\mathbf{d})\right) = \int_\Omega \psi\left(\mathbf{U}^{RK}(\mathbf{d})\right) d\Omega - F\left(\mathbf{U}^{RK}(\mathbf{d})\right) \tag{42}$$

In this stage, the smooth part of the solution by RK approximation is obtained. Note that this leads to a standard Galerkin-based RKPM formulation and the problem can be solved by using a standard matrix solver. Since the RK approximation obtained in the first stage can oscillate around the localization due to the Gibbs-type phenomenon, the filtered RK coefficients are used as the initial guess for the next step as follows:

$$\tilde{d}_I = \sum_{J=1}^{NP} \overline{\Psi}_J(\mathbf{x}_I) \bar{d}_J \tag{43}$$

where $\tilde{d}_I$ is the filtered RK coefficient at node $I$ and $\overline{\Psi}_J(\mathbf{x})$ is an RK filter[29]. As the purpose of filtering the RK coefficients is to provide a non-oscillatory initial guess to the next solution stage, many appropriate filters are possible. In this work, the RK shape function is used as the RK filter, that is, $\overline{\Psi}_J(\mathbf{x}) = \Psi_J(\mathbf{x})$.

B. <u>NN-RK Solution Stage</u>

In the second stage, Eq. (41) is solved for both $\mathbf{d}$ and $\mathbf{W}$. In the iterative neural network procedures, the initial guess of the RK coefficients is $\mathbf{d} = \tilde{\mathbf{d}}$. The neural network weights, $\mathbf{W} = \{\mathbf{W}^L, \mathbf{W}^S, \mathbf{W}^P\}$, are initialized as follows:

1) $\mathbf{W}^L$ is initialized such that the NN blocks are uniformly distributed over the domain,

2) $\mathbf{W}^S$ is initialized such that the kernels are uniformly distributed in the parametric coordinate in each of the NN blocks, and

3) $\mathbf{W}^P = \mathbf{0}$ is taken for the initiation of polynomial basis parameters.

The potential energy minimization problem is solved iteratively by a gradient descent-type optimizer. In this work, *Adam*[30], a first-order gradient-based stochastic optimizer with adaptive learning rate, is used. For the upper bound of the learning rate, the default value ($10^{-3}$) is initially used and the upper bound of the learning rate is decreased when severe oscillation in the loss curve is observed. The upper bounds used for the numerical examples in this work will be specified in Section 5.

# 5. Numerical Examples

A series of numerical examples is presented to demonstrate the localization capturing ability of the proposed method, including the automatic detection of localization locations and orientations, as well as the effectiveness of regularization. For the RK shape functions, the linear monomial basis functions and the cubic B-spline kernel function with normalized support size of 2.0 are used unless otherwise specified. For the domain integration, 2x2 Gauss quadrature is employed. For accurate numerical integration, the size of the integration cells is selected such that at least three integration cells are located within the localization width, and at least three integration cells are located between every two adjacent RK nodes. The distribution of the quadrature cells is described in each example. For the optimization in the NN enrichment solution, *Adam*, a first-order gradient-based stochastic optimizer is used. For the imposition of Dirichlet boundary conditions, a penalty number of $\alpha = 10^2$ in (1) is used.

## 5.1. One-dimensional elasticity with pre-degraded material

The following one-dimensional elastic bar problem is considered to investigate how the individual NN block approximations play a role in capturing localizations.

$$\min \Pi = \int_{-1}^{+1} [\frac{1}{2}E(x)u(x)_{,x}^2 - u(x)b(x)]\, dx \\ + \frac{1000E}{2h}[(u(-1) - g_1)^2 + (u(1) - g_2)^2] \tag{44}$$

where $E$, $b$, $h$, and $g_i$ are Young's modulus, body force, RK nodal spacing, and Dirichlet boundary value, respectively. The Dirichlet boundary values $g_1 = 0$ and $g_2 = 0.5$ are used. As shown in Figure 22, the material is pre-degraded locally with a small Young's modulus, and the bar is subjected to a smooth body force:

$$E(x) = E_0 \left[ 1.0 - \sum_{k=1}^{3} 0.99 \operatorname{sech}^2 \left( \frac{\max(0, |x - \bar{x}_k| - 0.002)}{0.01} \right) \right] \quad (45)$$

with $(\bar{x}_1, \bar{x}_2, \bar{x}_3) = (-0.65, -0.21, 0.55)$, and

$$b(x) = 10 \sin 3\pi x. \quad (46)$$

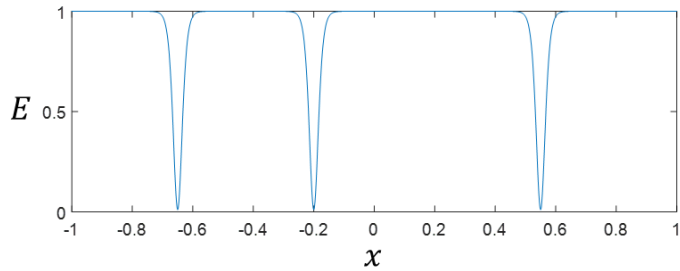

(a)

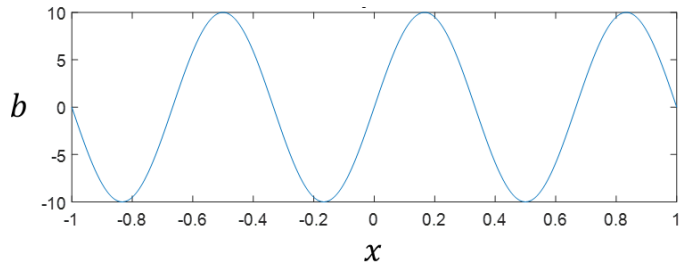

(b)

Figure 22. One-dimensional elasticity: (a) Young's modulus distribution and (b) body force distribution

The RK approximation space is constructed by 21 equally spaced RK points. A total of 500 uniformly distributed quadrature cells are employed, and the two-point Gauss quadrature is used for each cell. Four NN blocks ($N_B=4$) are initially uniformly distributed with quadratic monomial bases which amount to 36 total unknowns. The NN kernels with $N_K=2$ are uniformly distributed throughout the domain at the initial stage and the monomial coefficients are initialized as zero, which means that the neural network initially is not informed of the localizations. Note that the employment of only 21 RK nodes is purposely done as the sharp transition in the displacement is solely taken care of by the neural network and the RK approximation only targets the smooth part of the displacement.

Figure 23 shows the block-level approximation where each block locally influences the solution and captures nearby localizations. Figure 24 shows the total solution, NN approximation, and RK approximation. As expected, the NN approximation captures the very sharp solution transition while the RK approximation represents the overall smooth solution. As shown in Figure 25, the pure RK solution achieves a similar resolution with 801 unknowns while the transitions are insufficiently sharp even with 201 and 401 nodes, while the NN-enhanced approximation agrees very well with the exact solution with 57 unknowns, which is a 93% reduction in the number of unknowns. The number of unknowns required to capture localizations in multi-dimensional problems will be much more pronounced, and the proposed approach is expected to be even more effective compared to the standard RK approximation.

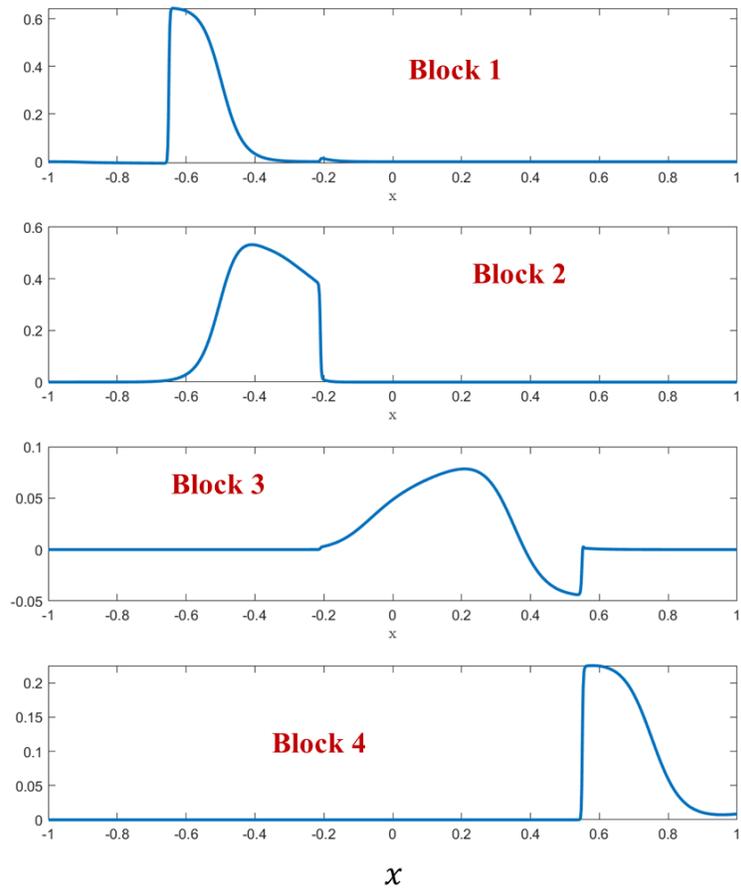

Figure 23. Block-level approximations

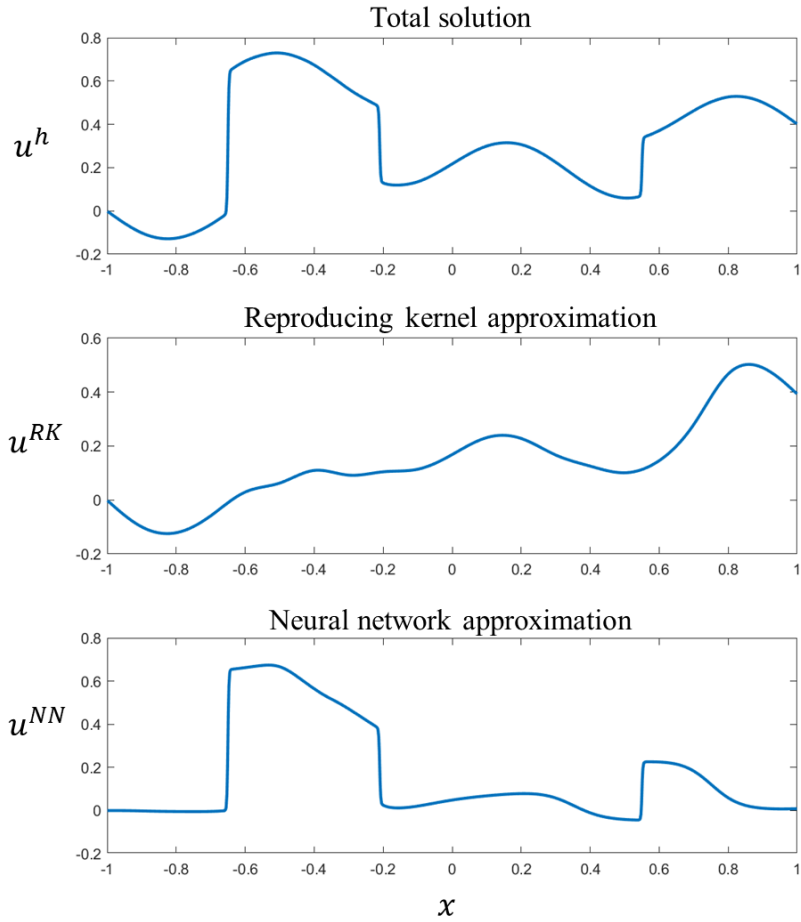

Figure 24. Numerical solution of one-dimensional problem: total solution, NN approximation, and RK approximation

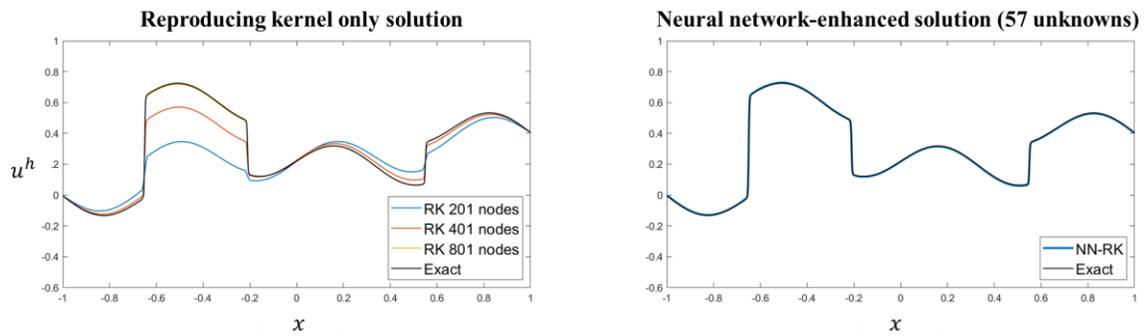

Figure 25. Pure RK solution (left) and NN-enhanced RK solution (right)

## 5.2. An elastic-damage bar under tension

To demonstrate the energy-based regularization effect along with the introduction of the NN length scale, a one-dimensional bar with imperfection under tension is considered. As shown in Figure 26, a one-dimensional bar in $\Omega = [-L/2, L/2]$ with a length of $L = 100.0$ and cross-sectional area of $A = 10.0$ is subjected to Dirichlet boundary conditions, $u(0) = -\bar{u}$ and $u(L) = \bar{u}$. A Young's Modulus of $E = 2.0 \times 10^6$ and fracture energy of $G_F = 1.885$ are taken[28]. Damage Model I, provided in Section 2.1, is used with the elastic strain limit of $\kappa_0 = 10^{-4}$ throughout the domain except for $x \in [-w/2, w/2]$ with $w = L/100$ in the middle of the bar where $k_0 = 0.95 \times 10^{-4}$ is used to initiate damage. The NN length scale, the lower bound of $c$ in Eq. (38), is selected as $\ell = w$ and $\kappa_c = 2G_F/(E\kappa_0 w)$ is used to ensure objective energy dissipation. Three RK discretizations with 11, 21, and 41 RK nodes are used in this study. For all cases, $N_K = 2$, and $N_B = 1$ are used with linear NN basis for the NN approximation, which involves 13 total NN parameters: 9 shape control parameters, and 4 monomial coefficients to be solved. For the domain integration, 2-point Gauss integration is used for 800 uniformly distributed integration domains.

The prescribed displacement $\bar{u}$ is gradually increased in three loading steps: 1) $\bar{u} = 0.95g$, 2) $\bar{u} = 1.4g$, and 3) $\bar{u} = 1.8g$ with $g = 0.5 \times 10^{-2}$. In the optimization procedure, 1000 epochs are used for the RK-only stages with an upper learning rate limit of $5 \times 10^{-7}$ applied to the *Adam* optimizer. For the NN-RK stage, 5000 epochs are used with the upper learning rate limits of $10^{-7}$ for the RK coefficients and the NN monomial coefficients and $10^{-5}$ for the other unknown NN parameters. Then, these learning rate limits are decreased by one third at every 1000 epochs thereafter.

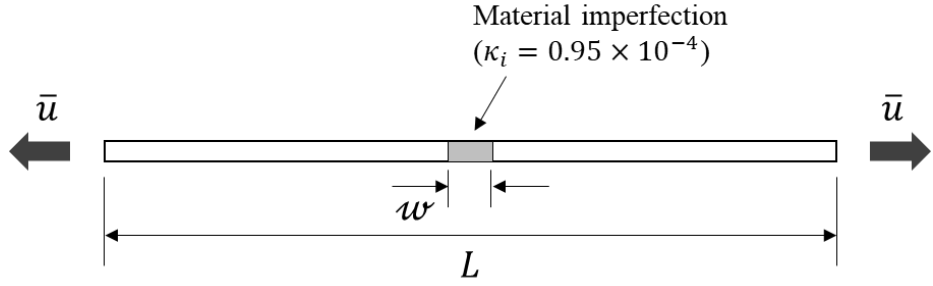

Figure 26. A one-dimensional bar with imperfection

Figure 27 (a) shows that all three models yield consistent load-displacement curves, which verified the regularization capability of the proposed method. The slope of the softening regime for a given fracture energy $G_F$ agrees very well with the reference solution obtained by Wei and Chen (2018)[28]. Contrarily, Figure 28 shows un-regularized load-displacement responses obtained in three cases with varying NN length scale proportional to the RK nodal spacing with the same damage model constants $\kappa_0$ and $k_c$ used. The localized strain field is shown in Figure 27 (b) where the strain distribution is converged as the domain is refined.

As shown in Figure 27 (c), the sudden jumps in the damage parameter, discussed in Section 3, are captured by the adaptive NN kernel functions. When smooth NN kernel functions with a fixed smoothness parameter $\beta = 5.0$ (see Section 3.2.1) is used without NN optimization, severely oscillating stress field around the localization point is obtained as shown in Figure 29 (a) while the NN kernel with adaptive $\beta$ via NN optimization yields a significant improvement as shown in Figure 29 (b). Note that Wei and Chen (2018)[28] bypassed this issue by using a gradient smoothing technique implemented with smooth RK kernel functions to obtain stress fields without oscillation. This shows potential of implementing a gradient smoothing technique to integrate the NN-enhanced formulation with smooth NN kernel functions, which will be a task of future work.

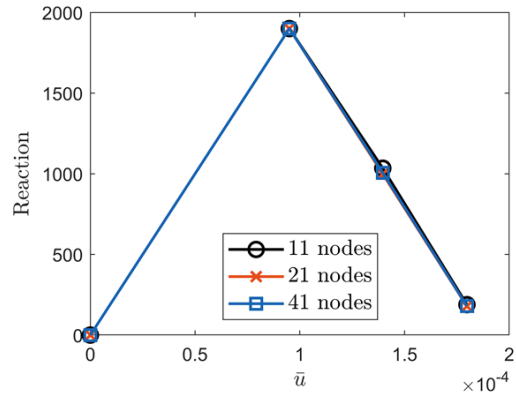

(a)

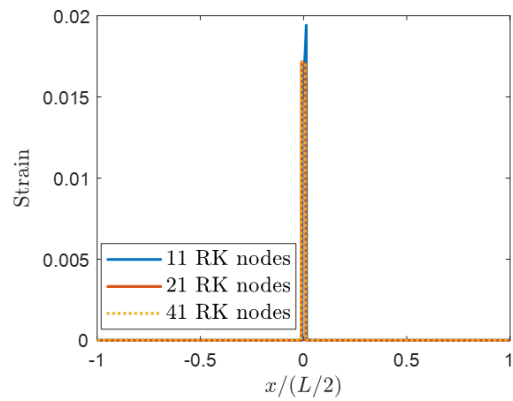

(b)

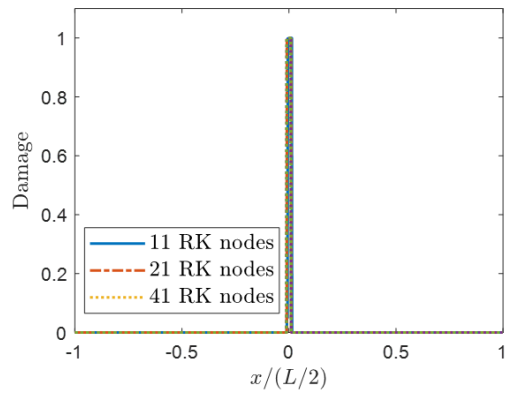

(c)

Figure 27. Solution material responses (1-D tensile bar) by the regularized model: (a) load-displacement curve, (b) strain, and (c) damage

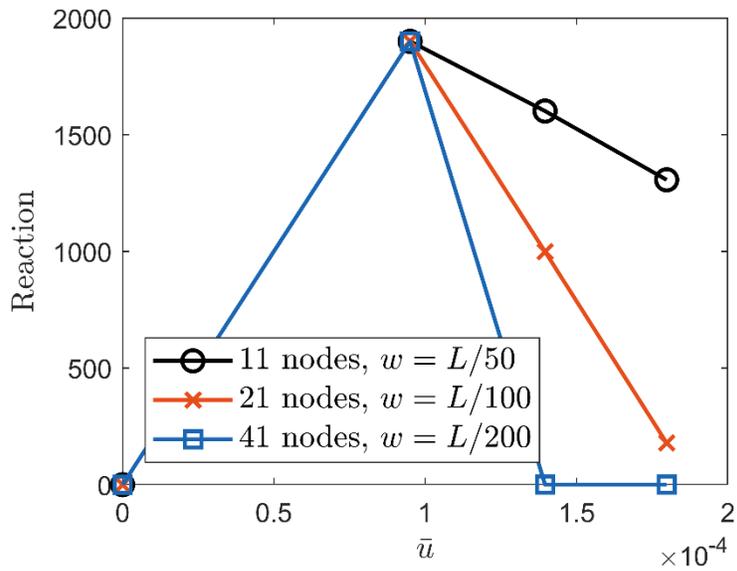

Figure 28. Load-displacement response of the un-regularized model

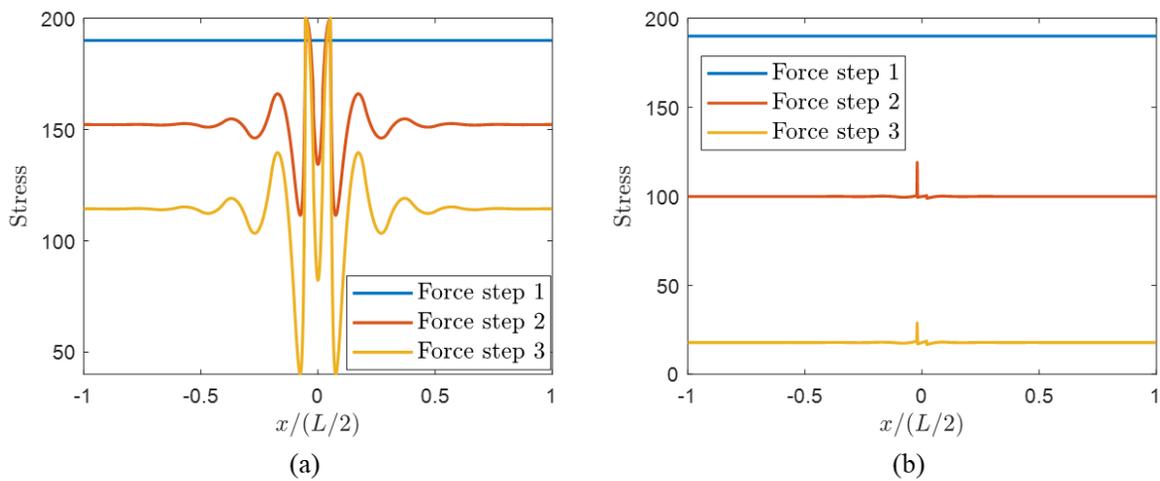

Figure 29. Stress fields obtained by (a) smooth NN kernel function with fixed parameter $\beta = 5.0$ in the NN optimization and (b) adaptive NN kernel function (21 RK nodes)

## 5.3. Tensile specimen with asymmetric imperfection

As shown in Figure 30, a tensile specimen with an asymmetric imperfection is considered. The material properties used in Section 5.2 are employed for the entire domain $\Omega = [0, H] \times [-H, H]$ where $H = 30$ and the material point $x \in [0, H/32] \times [-H/32, H/32]$ is initially weakened with $\kappa_0 = 0.99 \times 10^{-4}$. The NN length scale of $\ell = H/32$ is used. Three RK discretizations with 7×7, 13×13, and 25×25 RK nodes are used for this study. For all cases, $N_{NR} = 16$, $N_K = 4$, and $N_B = 1$ are used with linear NN basis for the NN approximation, which involves 124 total unknown NN parameters: 82 parameterization weights, 18 shape control parameters, and 24 monomial coefficients. Due to the symmetry of the problem, only the upper half of the domain is explicitly modeled and, for the domain integration, 2×2 Gauss integration is used for 96x96 uniformly distributed integration cells.

The prescribed displacement $\bar{u}$ is gradually increased in five loading steps: 1) $\bar{u} = g$, 2) $\bar{u} = 1.5g$, 3) $\bar{u} = 2.0g$, 4) $\bar{u} = 2.5g$, and 5) $\bar{u} = 3.0g$ where $g = 2.76 \times 10^{-3}$. In the optimization procedure, 2000 epochs are used for the RK-only stages with an upper learning rate limit of $10^{-7}$ applied to the *Adam* optimizer. For the NN-RK stage, 15000 epochs are used for the first loading step and 5000 epochs are used for the other loading steps for all models. For the upper learning rate limits, $10^{-7}$ is used to obtain the RK coefficients and the NN monomial coefficients and $10^{-5}$ is used to obtain the other unknown NN parameters.

As shown in Figure 31, the proposed method yields consistent damage patterns on the four discretizations, which agree well with the damage pattern obtained by the implicit gradient RK regularization in Chen et al. (2000)[31]. The force-displacement curves are plotted in Figure 32 where all three models result in nearly identical reaction force profiles, which demonstrates the regularization effect of the proposed approach in multi-dimensional problems. Note that in the three employed RK discretization models, the RK nodal spacings are larger than the NN length scales, thus the sharp displacement transition due to the strain localization is entirely captured by the NN approximation (see Figure 33).

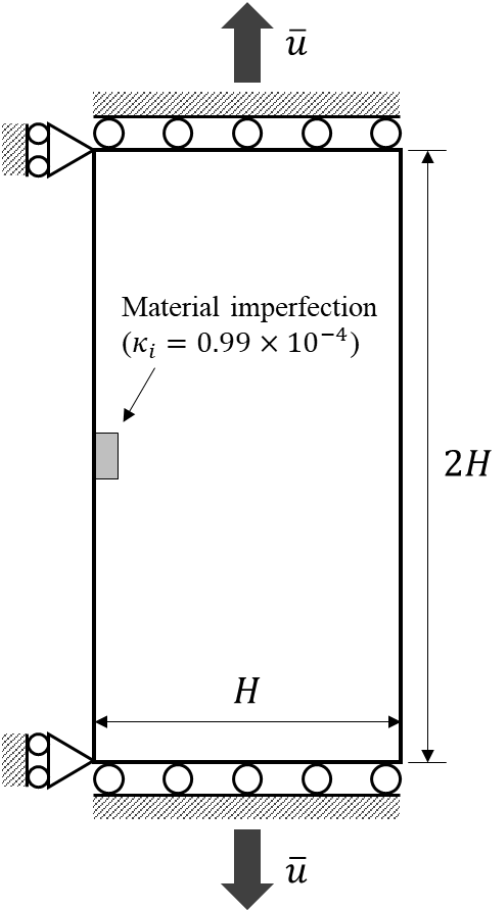

Figure 30. A 2-dimensional tensile specimen with asymmetric imperfection

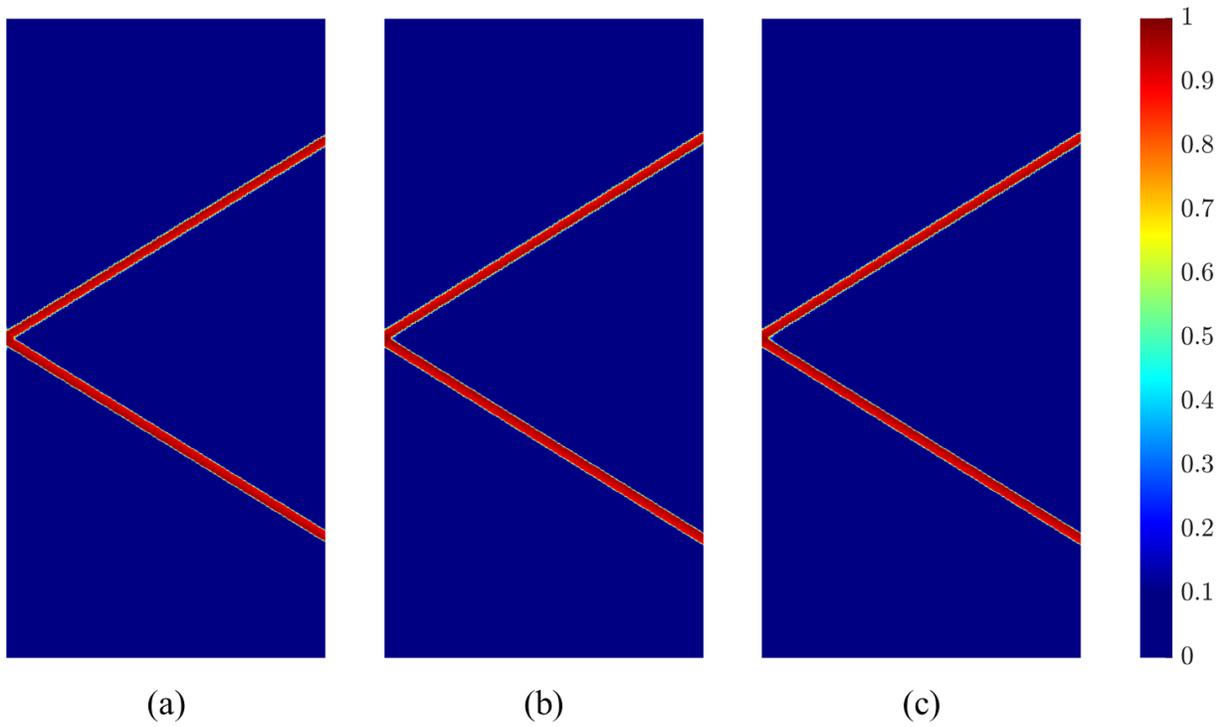

Figure 31. Damage pattern on 2-D tensile specimen: (a) 7x7 Nodes, (b) 13x13 Nodes, (c) 25x25 Nodes

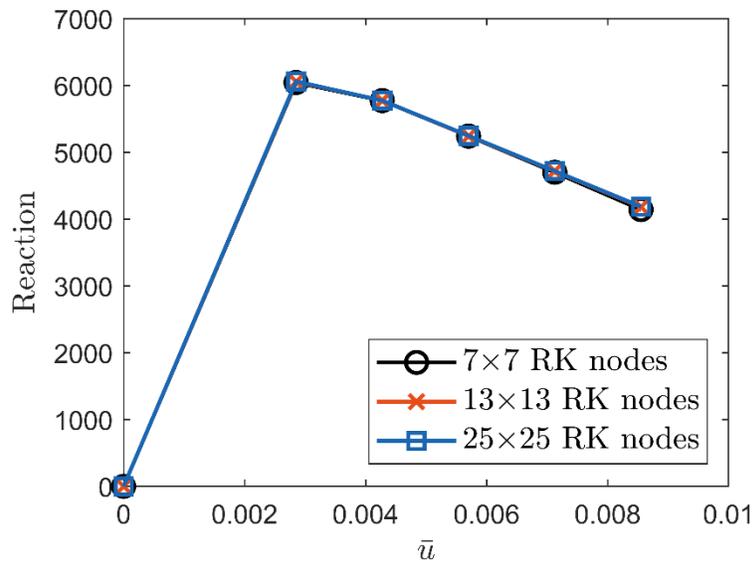

Figure 32. Regularized force-displacement curve (2-D tensile specimen)

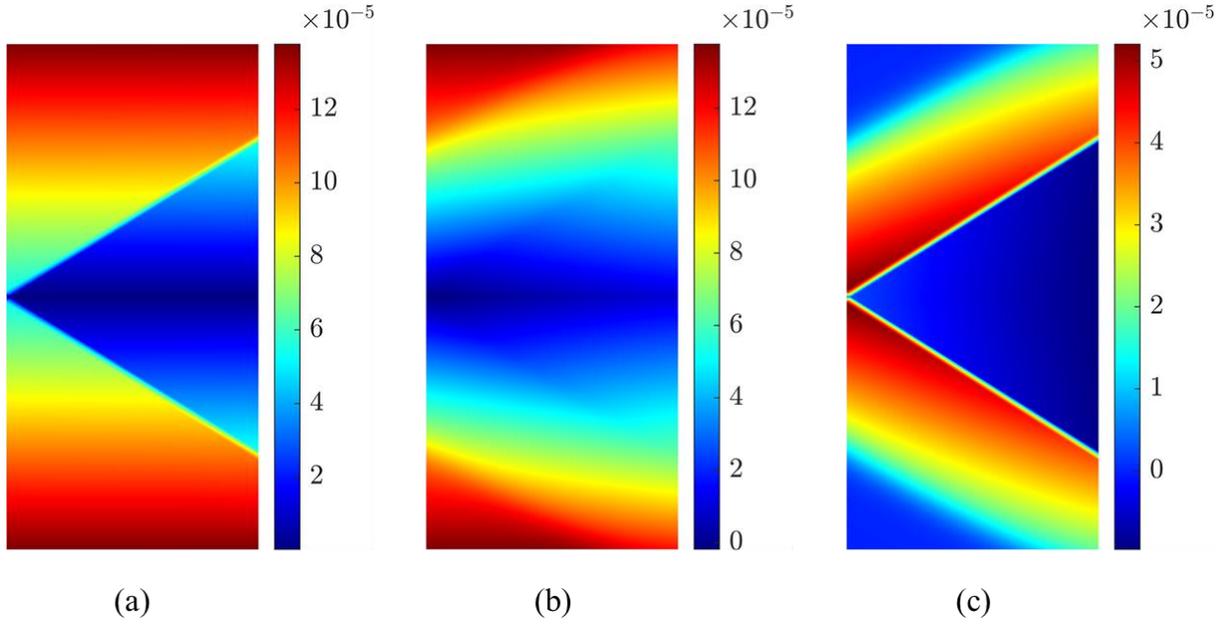

Figure 33. Vertical displacement of M3-3: (a) $u_2^h$, (b) $u_2^{RK}$, (c) $u_2^{NN}$

## 5.4. Pre-notched specimen under simple shear

To demonstrate the capability of the proposed method to model evolving localization paths, a pre-notched specimen under simple shear, as shown in Figure 34, is considered. In this work, the following damage model is applied:

$$\eta(\kappa) = \frac{2\kappa}{2\kappa + \bar{p}} \tag{47}$$

$$g(\eta) = (1 - \eta)^2 \tag{48}$$

$$\bar{\psi}(\eta) = \frac{\bar{p}}{2}\eta^2 \qquad (49)$$

For Kuhn-Tucker conditions in (8), $\bar{\kappa} = \psi_0^+$ is used. This damage model is derived from the phase field formulation provided in Miehe et al. (2010)[13] in the absence of the higher order phase field term, which served as a regularization in the phase field model. In this work, regularization in the NN approximation proposed in Section 3.3 is employed. A Young's modulus of 210 GPa, Poisson's ratio of 0.3, and $G_F$=2.7 N/mm are chosen[13,16], and the NN length scale, $\ell$=0.0175 mm, and $\bar{p} = 2G_F/\ell$ are used. Material is pre-degraded by pre-existing damage field as shown in Figure 34(b). The domain is uniformly discretized by 256 RK nodes for the RK approximation, and four NN blocks are used with cubic basis for the NN approximation. The total number of unknowns is 790, including 512 RK unknowns and 278 NN unknowns. The Gauss quadrature points used for this simulation are plotted in Figure 34(c). Figure 35 shows the evolution of damage produced with respect to the shear displacement $g$. The series of results demonstrates that the proposed method is capable of capturing the evolution of localizations, including the damage evolution and the sharp transition in $u_1^h$ and $u_2^h$ across the localization, which is a promising result. The load-displacement curve shown in Figure 36 as well as the angle of crack path of 63° degrees compares well with the initial angle of 65° reported in Miehe et al. (2010)[13].

The total CPU time of 572 minutes was used by the proposed method with a single GPU, NVIDIA A100 with 40 GB memory. In literature[32], it was reported that CPU times of 1304 min and 957 min were required by a traditional phase field approach with 250×250 standard finite elements and a phase field method with an adaptive refinement, respectively. This comparison is encouraging although different program languages and hardware were employed. Also, it is worth noting that, in the proposed method, the NN approximation space is constructed based on a highly sparse neural network with interpretable NN weights and biases. As the computational efficiency that can be gained by utilizing sparse matrices becomes more significant with a larger problems size compared to using full matrices, it is expected that the CPU time saving achieved by the proposed method will also be more significant in solving larger scale problems.

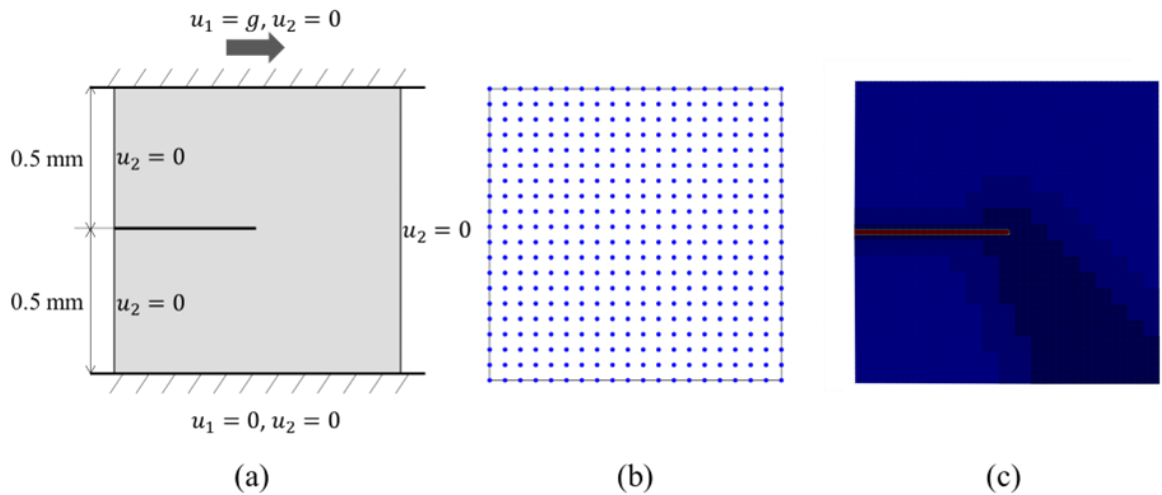

Figure 34. Problem setting of elastic-damage problem: (a) geometry and boundary conditions, (b) RK discretization, and (c) pre-existing damage and Gauss integration points

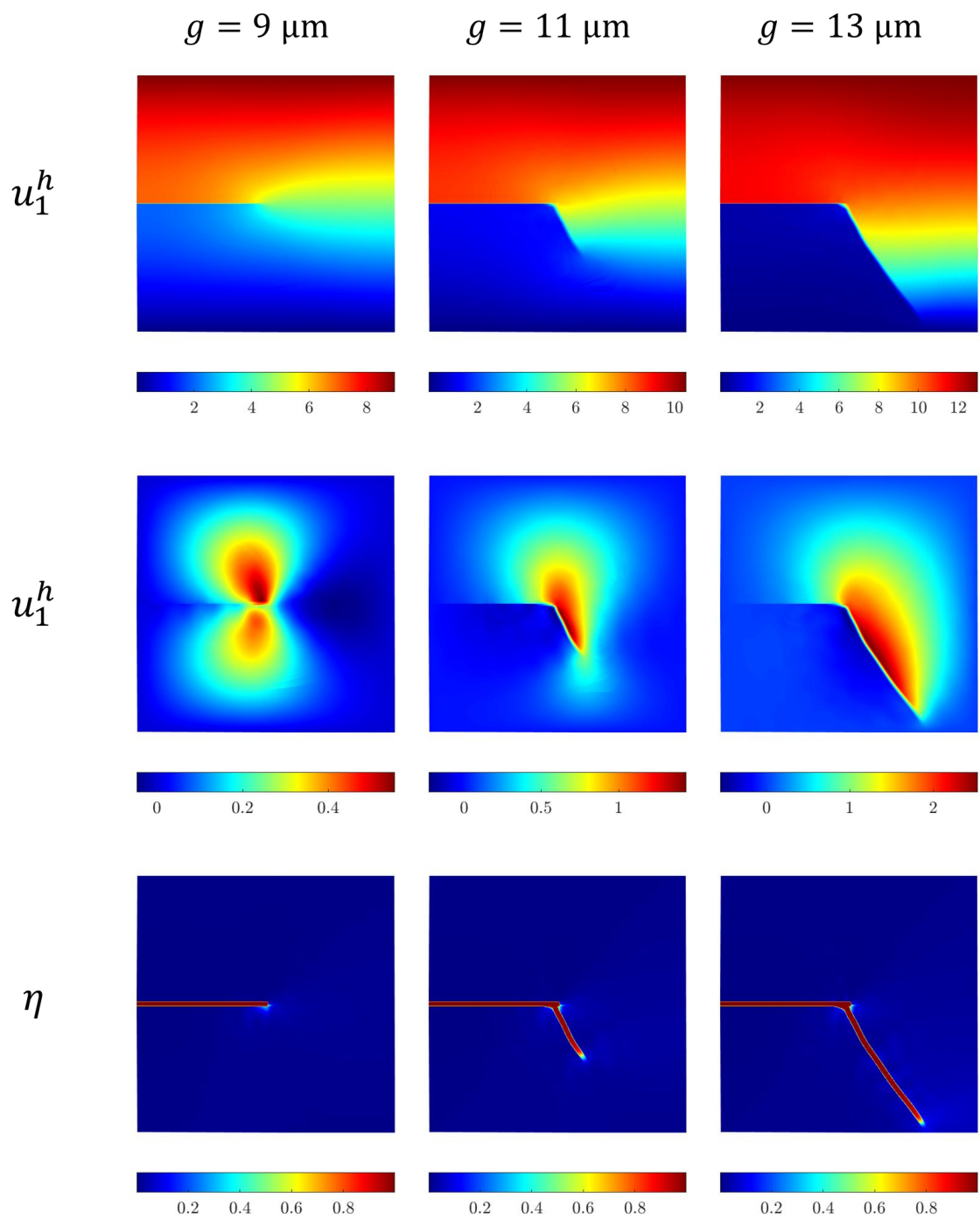

Figure 35. Damage evolution: (left) $g = 9 \times 10^{-3}$ mm, (center) $g = 11 \times 10^{-3}$ mm, (right) $g = 13 \times 10^{-3}$ mm; (top) $u_1^h$, (middle) $u_2^h$, (bottom) damage

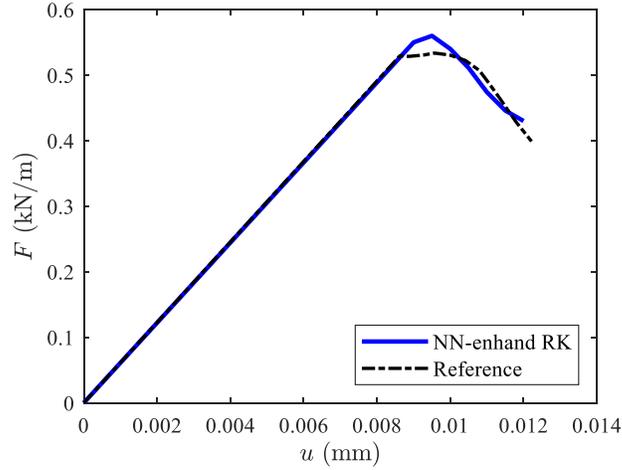

Figure 36. Load-displacement curve

## 5.5. L-Shaped Panel Test

To simulate relatively complex crack propagations, an L-shaped panel test[33] is explored. The geometry and boundary conditions, summarized in Figure 37(a), illustrate a fixed bottom-left boundary and an applied upward displacement $g$ on the horizontal arm of the L-shaped panel. Material parameters selected for validation purposes are Young's modulus of $E = 16.0$ kN/mm$^2$, Poisson's ratio of $\nu = 0.18$, the fracture energy of $G_F = 65$ N/m, the damage model parameter $\kappa_0 = 2.34 \times 10^{-4}$, and the NN length scale of $\ell = 3.125$ mm along with the damage model provided in Eqs. (5)-(7). Using Eq. (40), the damage parameter $\kappa_c = 2G_F/(E\kappa_0\ell)$ is determined for an objective energy dissipation. Following Unger et al. (2007)[34], the material parameters given by Winkler (2001)[33] are adjusted such that the load-displacement response agrees with the experimental data. The domain is uniformly discretized by 341 RK nodes for the RK approximation as shown in Figure 37(b). For the NN approximation, a single NN block with $N_{NR} = 32$, $N_K = 4$, and $N_B = 1$ is used. The total number of unknowns is 899, including 682 RK unknowns and 217 NN unknowns. The Gauss quadrature points used for this simulation are shown in Figure 37(c). The numerically predicted damage fields are shown in Figure 38 where localization is initiated in a diagonal direction and changes its direction to the horizontal as the vertical deflection increases. As shown in Figure 38(d), the predicted damage path lies within the experimentally observed crack path range[33]. Figure 39 shows the load-displacement curve predicted by the proposed method with a good agreement with the experimental data[33].

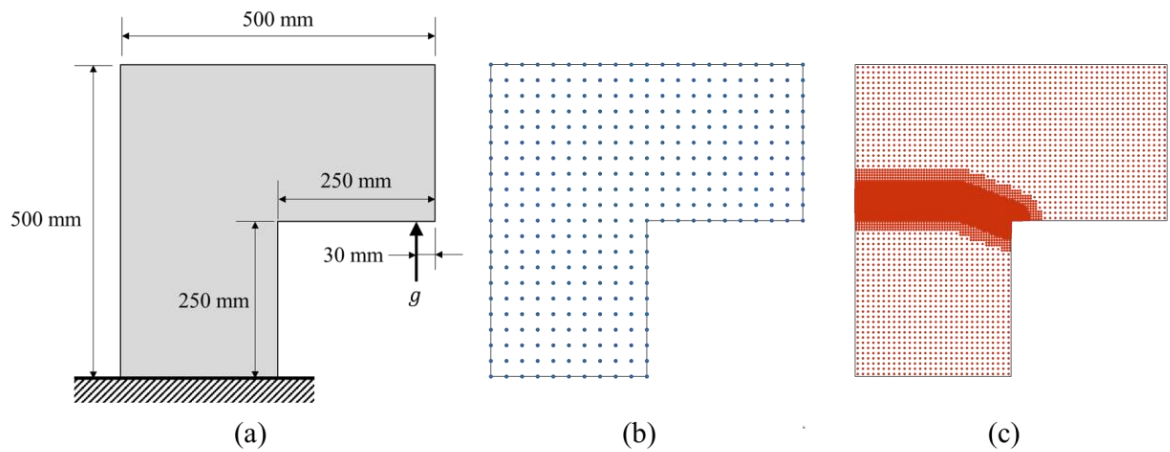

Figure 37. L-shaped Panel Simulation Setup: (a) problem geometry and boundary conditions, (b) RK discretization, and (c) Gauss integration points

Figure 38. Damage field at (a) $g$=0.20 mm, (b) $g$=0.28 mm, (c) $g$=0.36 mm, and (d) comparison with the experimentally observed crack range indicated by the white curve[33]

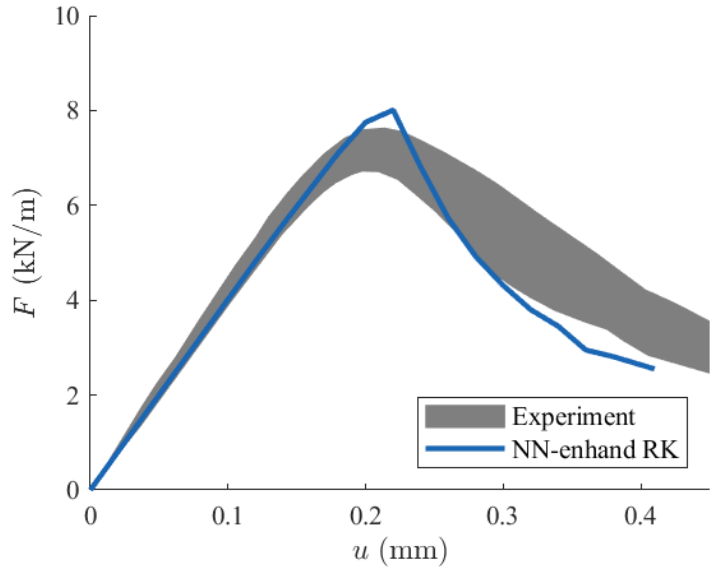

Figure 39. Load-displacement curves showing comparison of NN-enhanced RK with experimental[33] data.

## 6. Conclusion

A neural network-enhanced reproducing kernel particle method has been proposed for modeling strain localization. The approximation space is constructed by enriching a standard Reproducing Kernel (RK) approximation with a neural network (NN) approximation. In the proposed function space, the NN approximation is constructed to capture the sharp solution transition near the region of strain localization while the RK approximation is utilized to approximate the smooth part of the solution. The capability of the RK approximation to accurately represent smooth functions with a coarse discretization allows the use of a small number of fixed RK shape functions for computational efficiency. In the NN approximation, adaptive NN kernel functions defined in a parametric coordinate introduce localized solution gradients with various types of topological patterns. The localization shape of the NN kernel functions is controlled by unknown shape parameters that are determined by the NN optimization. The parametrization is performed by the blocked parametrization network, with the weights of the network controlling the location and orientation of the localization, and they are also determined by the optimization. One four-kernel NN block is capable of capturing a triple junction or a quadruple junction topological pattern. More complex topology can be captured by the superposition of multiple block-level NN approximations, each of which is defined in its own parametric domain. A series of function approximation testing problems showed that 1) the numerical solution converges with an increased number of neurons in the parameterization network and 2) the number of four-kernel NN blocks needed to capture complex localization topology with multiple triple junctions is greater than or equal to the number of the triple junctions.

To achieve a discretization-insensitive solution in modeling localization, a regularization of the NN-enhanced RK approximation has been introduced such that the localization width of NN kernels is determined by the NN length scale parameter, which leads to the dissipation energy objectivity. A set of numerical examples have been analyzed to demonstrate that the proposed method is effective in modeling evolving localization paths with discretization-insensitive solutions.

## Acknowledgments

The supports of this work by the Sandia National Laboratories to UC San Diego under Contract Agreement 1655264 is greatly acknowledged. Sandia National Laboratories is a multi-mission laboratory managed and operated by National Technology and Engineering Solutions of Sandia, LLC, a wholly owned subsidiary of Honeywell International Inc., for the U.S. Department of Energy's National Nuclear Security Administration under contract DE-NA0003525. This paper describes objective technical results and analysis. Any subjective views or opinions that might be expressed in the paper do not necessarily represent the views of the U.S. Department of Energy or the United States Government.

# Appendix A. Target Function Construction

The expression used to construct target function TF3 used for the convergence study in Section 3.2.4 is provided. As shown in Figure 40, TF3 decomposes the domain into five constant-valued sections separated by transition zones, which are defined by intersecting circular arcs and the domain boundaries. A total of three triple-junctions are formed using seven arc transition zones, each of constant width $w = 0.04$. Each arc is defined using two vertex locations and the arc radius. All eight vertices are positioned at [(-1,-0.1), (-0.3,0.2), (0.3,0.6), (0.2,-0.5), (0.4,1), (1,0.3), (1,-0.4), (0.2,-1)], and the seven arc radii used are [3, 2, 2, 1.5, 1.5, 1.5, 1.5]. TF3 is constructed through the summation of five functions, each of which has a constant value within one of the five subdomains and zero outside, with a ramp in the transition zone. The explicit expression of TF3 is as follows:

$$f^{TF3}(\mathbf{x}) = \sum_{K=1}^{N_{section}} \bar{f}_K g_K(\mathbf{x}) \tag{50}$$

$$g_K(\mathbf{x}) = \prod_{\beta=1}^{N_{arc}^K} \left( \frac{1}{2} + \frac{s_{K\beta}}{2} h_{K\beta}(\mathbf{x}) \right) \tag{51}$$

$$h_{K\beta}(\mathbf{x}) = \begin{cases} -1 & \|\mathbf{x} - \mathbf{c}_{K\beta}\| \leq R_{K\beta} - \bar{\ell}/2 \\ 1 & \|\mathbf{x} - \mathbf{c}_{K\beta}\| \leq R_{K\beta} + \bar{\ell}/2 \\ \dfrac{\|\mathbf{x} - \mathbf{c}_{K\beta}\| - R_{K\beta}}{\bar{\ell}} & \text{otherwise} \end{cases} \tag{52}$$

where $N_{section}$ and $N_{arc}$ are the number of sections and the number of arcs of Section $K$, respectively. The definitions used in Eqs. (50) - (52) are provided below:

$f^{TF3}(\mathbf{x})$: target function TF3

$\bar{f}_K$: Value $f^{TF3}(\mathbf{x})$ in section $K$

$g_K(\mathbf{x})$: Subdomain definition for Section K. This has a value of 1 within the section, 0 outside, and between 0 and 1 in the transition zone.

$s_{K\beta}$: sign coefficient of arc $\beta$, section $K$. -1 if the arc is convex outward from section $K$ and 1 if the arc is convex inward to section $K$.

$h_{K\beta}$: piece-wise step function

$\bar{\ell}$: width of the transition zone.

$R_{K\beta}$: radius of arc $\beta$, section $K$

$\mathbf{c}_{K\beta}$: center of arc $\beta$, section $K$

$\|\cdot\|$: Euclidean norm

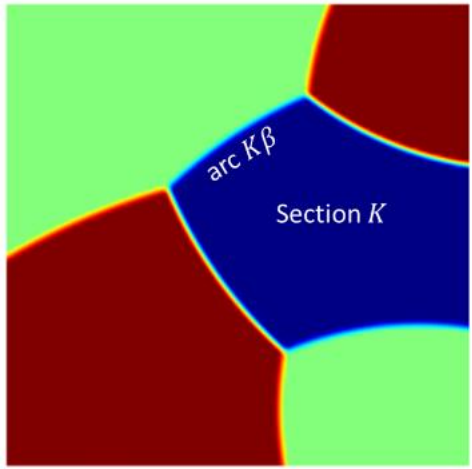

Figure 40. TF3 schematic for definition of Section K